\documentclass[twocolumn]{aastex63}

\usepackage{amsmath}
\usepackage{lineno}

\newcommand{\planck}{{\sl Planck}}

\definecolor{amethyst}{rgb}{0.6, 0.4, 0.8}
\definecolor{byzantine}{rgb}{0.74, 0.2, 0.64}

\newcommand{\Tcmb}{\mbox{$T_{\mbox{\tiny CMB}}$}}
\newcommand{\muk}{\ensuremath{\mu {\rm K}}}

\newcommand{\sqdeg}{deg$^2$}
\newcommand{\um}{\ensuremath{\mu{\rm m}} }

\newcommand{\webaddress}{\url{http://pole.uchicago.edu/public/data/sptsz_ymap}}
\newcommand{\webaddresslambda}{\url{https://lambda.gsfc.nasa.gov/product/spt/spt_prod_table.cfm}}

\newcommand{\fsz}{\ensuremath{f_\mathrm{\mbox{\tiny{SZ}}}}}
\newcommand{\ysz}{\ensuremath{y_\mathrm{\mbox{\tiny{SZ}}}}}
\newcommand{\tcmb}{\ensuremath{T_\mathrm{\mbox{\tiny{CMB}}}}}

\newcommand{\beq}{\begin{equation}}
\newcommand{\eeq}{\end{equation}}
\newcommand{\beqar}{\begin{eqnarray}}
\newcommand{\eeqar}{\end{eqnarray}}

\graphicspath{{FIGURES/}}

\begin{document}

\title{CMB/kSZ and Compton-$y$ Maps from 2500 square degrees of SPT-SZ and \planck \ Survey Data}


\newcommand{\McGill}{Department of Physics and McGill Space Institute, McGill University, Montreal, Quebec H3A 2T8, Canada}
\newcommand{\KIPAC}{Kavli Institute for Particle Astrophysics and Cosmology, Stanford University, 452 Lomita Mall, Stanford, CA 94305}
\newcommand{\Stanford}{Dept. of Physics, Stanford University, 382 Via Pueblo Mall, Stanford, CA 94305}
\newcommand{\Davis}{Department of Physics and Astronomy, University of California, Davis, CA, USA 95616}
\newcommand{\Penn}{Center for Particle Cosmology, Department of Physics and Astronomy, University of Pennsylvania, Philadelphia, PA,  USA 19104} 
\newcommand{\KICPChicago}{Kavli Institute for Cosmological Physics, University of Chicago, Chicago, IL, USA 60637}
\newcommand{\PhysicsUChicago}{Department of Physics, University of Chicago, Chicago, IL, USA 60637}
\newcommand{\AAUChicago}{Department of Astronomy and Astrophysics, University of Chicago, Chicago, IL, USA 60637}
\newcommand{\FNAL}{Fermi National Accelerator Laboratory, MS209, P.O. Box 500, Batavia, IL 60510}
\newcommand{\ArgonneHEP}{High Energy Physics Division, Argonne National Laboratory, Argonne, IL, USA 60439}
\newcommand{\EFIChicago}{Enrico Fermi Institute, University of Chicago, Chicago, IL, USA 60637}
\newcommand{\SLAC}{SLAC National Accelerator Laboratory, 2575 Sand Hill Road, Menlo Park, CA 94025}
\newcommand{\Caltech}{California Institute of Technology, Pasadena, CA, USA 91125}
\newcommand{\Berkeley}{Department of Physics, University of California, Berkeley, CA, USA 94720}
\newcommand{\Cifar}{Canadian Institute for Advanced Research, CIFAR Program in Cosmology and Gravity, Toronto, ON, M5G 1Z8, Canada}
\newcommand{\Colorado}{Center for Astrophysics and Space Astronomy, Department of Astrophysical and Planetary Sciences, University of Colorado, Boulder, CO, 80309}
\newcommand{\ESO}{European Southern Observatory, Karl-Schwarzschild-Stra{\ss}e 2, 85748 Garching, Germany}
\newcommand{\Colphys}{Department of Physics, University of Colorado, Boulder, CO, 80309}
\newcommand{\Illast}{Astronomy Department, University of Illinois at Urbana-Champaign, 1002 W. Green Street, Urbana, IL 61801, USA}
\newcommand{\Illphys}{Department of Physics, University of Illinois Urbana-Champaign, 1110 W. Green Street, Urbana, IL 61801, USA}
\newcommand{\UChicago}{University of Chicago, Chicago, IL, USA 60637}
\newcommand{\LBNL}{Physics Division, Lawrence Berkeley National Laboratory, Berkeley, CA, USA 94720}
\newcommand{\Mcmaster}{Department of Physics and Astronomy, McMaster University, 1280 Main St. W., Hamilton, ON L8S 4L8, Canada}
\newcommand{\Michigan}{Department of Physics, University of Michigan, Ann  Arbor, MI, USA 48109}
\newcommand{\Munich}{Faculty of Physics, Ludwig-Maximilians-Universit\"{a}t, 81679 M\"{u}nchen, Germany}
\newcommand{\ExcellenceCluster}{Excellence Cluster ORIGINS, Boltzmannstr. 2, 85748 Garching, Germany}
\newcommand{\MPE}{Max-Planck-Institut f\"{u}r extraterrestrische Physik, 85748 Garching, Germany}
\newcommand{\Dunlap}{Dunlap Institute for Astronomy \& Astrophysics, University of Toronto, 50 St George St, Toronto, ON, M5S 3H4, Canada}
\newcommand{\Minnesota}{Department of Physics, University of Minnesota, Minneapolis, MN, USA 55455}
\newcommand{\Melbourne}{School of Physics, University of Melbourne, Parkville, VIC 3010, Australia}
\newcommand{\CaseWestern}{Physics Department, Center for Education and Research in Cosmology and Astrophysics, Case Western Reserve University,Cleveland, OH, USA 44106}
\newcommand{\ArtInstChicago}{Liberal Arts Department, School of the Art Institute of Chicago, Chicago, IL, USA 60603}
\newcommand{\JPL}{Jet Propulsion Laboratory, California Institute of Technology, Pasadena, CA 91109, USA}
\newcommand{\CfA}{Center for Astrophysics $|$ Harvard \& Smithsonian, 60 Garden Street, Cambridge, MA 02138, USA}
\newcommand{\UToronto}{Department of Astronomy \& Astrophysics, University of Toronto, 50 St George St, Toronto, ON, M5S 3H4, Canada}
\newcommand{\BCCP}{Berkeley Center for Cosmological Physics, Department of Physics, University of California, and Lawrence Berkeley National Labs, Berkeley, CA, USA 94720}
\newcommand{\Arizona}{Steward Observatory, University of Arizona, 933 North Cherry Avenue, Tucson, AZ 85721, USA}
\newcommand{\Toronto}{Department of Astronomy \& Astrophysics, University of Toronto, 50 St George St, Toronto, ON, M5S 3H4, Canada}
\newcommand{\UCLA}{Department of Physics and Astronomy, University of California, Los Angeles, CA 90095, USA}
\newcommand{\MichSt}{Department of Physics and Astronomy, Michigan State University, East Lansing, MI 48824, USA}
\newcommand{\KEK}{High Energy Accelerator Research Organization (KEK), Tsukuba, Ibaraki 305-0801, Japan}


\author[0000-0001-7665-5079]{L.~E.~Bleem}
\affiliation{\ArgonneHEP} 
\affiliation{\KICPChicago}

\author[0000-0001-9000-5013]{ T.~M.~Crawford}
\affiliation{\KICPChicago}
\affiliation{\AAUChicago}

\author[0000-0002-6443-3396]{B.~Ansarinejad}
\affiliation{\Melbourne}

\author[0000-0002-5108-6823]{B.~A.~Benson}
\affiliation{\FNAL}
\affiliation{\KICPChicago}
\affiliation{\AAUChicago}

\author[0000-0002-4900-805X]{ S.~Bocquet}
\affiliation{\Munich}
\affiliation{\ExcellenceCluster}

\author[0000-0002-2044-7665]{J.~E.~Carlstrom}
\affiliation{\KICPChicago}
\affiliation{\AAUChicago}
\affiliation{\PhysicsUChicago}
\affiliation{\ArgonneHEP}
\affiliation{\EFIChicago}

\author[0000-0002-6311-0448]{ C.~L.~Chang}
\affiliation{\ArgonneHEP}
\affiliation{\KICPChicago}
\affiliation{\AAUChicago}

\author[0000-0001-8241-7704]{ R.~Chown}
\affiliation{\Mcmaster}

\author{ A.~T.~Crites}
\affiliation{\Toronto}
\affiliation{\KICPChicago}
\affiliation{\AAUChicago}

\author{ T.~de~Haan}
\affiliation{\KEK}
\affiliation{\Berkeley}

\author{ M.~A.~Dobbs}
\affiliation{\McGill}
\affiliation{\Cifar}

\author [0000-0002-5370-6651] {W.~B.~Everett}
\affiliation{\Colorado}

\author[0000-0001-7874-0445]{ E.~M.~George}
\affiliation{\ESO}
\affiliation{\Berkeley}

\author{R.~Gualtieri}
\affiliation{\ArgonneHEP} 

\author[0000-0003-2606-9340]{ N.~W.~Halverson}
\affiliation{\Colorado}
\affiliation{\Colphys}

\author{ G.~P.~Holder}
\affiliation{\Illast}
\affiliation{\Illphys}

\author{ W.~L.~Holzapfel}
\affiliation{\Berkeley}

\author{ J.~D.~Hrubes}
\affiliation{\UChicago}

\author{ L.~Knox}
\affiliation{\Davis}

\author[0000-0003-3106-3218]{ A.~T.~Lee}
\affiliation{\Berkeley}
\affiliation{\LBNL}

\author{ D.~Luong-Van}
\affiliation{\UChicago}

\author[0000-0002-2367-1080]{ D.~P.~Marrone}
\affiliation{\Arizona}

\author{ J.~J.~McMahon}
\affiliation{\KICPChicago}
\affiliation{\AAUChicago}
\affiliation{\PhysicsUChicago}
\affiliation{\EFIChicago}

\author{ S.~S.~Meyer}
\affiliation{\KICPChicago}
\affiliation{\AAUChicago}
\affiliation{\PhysicsUChicago}
\affiliation{\EFIChicago}

\author{ M.~Millea}
\affiliation{\Berkeley}

\author{ L.~M.~Mocanu}
\affiliation{\KICPChicago}
\affiliation{\AAUChicago}

\author[0000-0002-6875-2087]{ J.~J.~Mohr}
\affiliation{\Munich}
\affiliation{\ExcellenceCluster}
\affiliation{\MPE}

\author{ T.~Natoli}
\affiliation{\KICPChicago}
\affiliation{\AAUChicago}

\author{Y.~Omori}
\affiliation{\KICPChicago}
\affiliation{\AAUChicago}
\affiliation{\KIPAC}
\affiliation{\Stanford}

\author{ S.~Padin}
\affiliation{\Caltech}
\affiliation{\KICPChicago}
\affiliation{\AAUChicago}

\author{ C.~Pryke}
\affiliation{\Minnesota}

\author{S.~Raghunathan}
\affiliation{\UCLA}
\affiliation{\MichSt}

\author[0000-0003-2226-9169]{ C.~L.~Reichardt}
\affiliation{\Melbourne}

\author{ J.~E.~Ruhl}
\affiliation{\CaseWestern}

\author{ K.~K.~Schaffer}
\affiliation{\ArtInstChicago}
\affiliation{\KICPChicago}
\affiliation{\EFIChicago}

\author[0000-0002-2757-1423]{ E.~Shirokoff}
\affiliation{\KICPChicago} 
\affiliation{\AAUChicago} 
\affiliation{\Berkeley} 

\author{ Z.~Staniszewski}
\affiliation{\JPL}
\affiliation{\CaseWestern}

\author[0000-0002-2718-9996]{ A.~A.~Stark}
\affiliation{\CfA}

\author[0000-0001-7192-3871]{J.~D.~Vieira}
\affiliation{\Illast} 
\affiliation{\Illphys} 

\author{R.~Williamson}
\affiliation{\JPL}
\affiliation{\KICPChicago} 
\affiliation{\AAUChicago}

\shortauthors{L. Bleem, T. Crawford, et al.}  

\hfill \break

\begin{abstract}
We present component-separated maps of the primary cosmic microwave background/kinematic Sunyaev-Zel'dovich (SZ) amplitude
and the thermal SZ Compton-$y$ parameter, created using data from the South Pole Telescope (SPT) and the \planck\ satellite.
These maps, which cover the $\sim$2500 square degrees of the Southern sky imaged by the SPT-SZ survey, represent a significant improvement over previous such products available in this region by virtue of their higher angular resolution ($1\farcm{25}$ for our highest-resolution Compton-$y$ maps) and lower noise at small angular scales. 
In this work we detail the construction of these maps using linear combination techniques, including our method for limiting the correlation of our lowest-noise Compton-$y$ map products with the cosmic infrared background. 
We perform a range of validation tests on these data products to test our sky modeling and combination algorithms, 
and we find good performance in all of these tests. 
Recognizing the potential utility of these data products for a wide range of astrophysical and cosmological analyses, 
including studies of the gas properties of galaxies, groups, and clusters, we make these products publicly available at \webaddress \ and on the NASA/LAMBDA website.  
\end{abstract}


\correspondingauthor{L. Bleem}
\email{lbleem@anl.gov}

\section{Introduction} \label{sec:intro}

Modern arcminute-scale resolution experiments such as the Atacama Cosmology Telescope \citep[ACT;][]{swetz11}, \planck \ \citep{planck11-1}, and the South Pole Telescope \citep[SPT;][]{carlstrom11} have been used to make exquisite temperature and polarization maps of the millimeter-wave sky. 
While the dominant signal in these observations arises from the primary cosmic microwave background (CMB), there is a wealth of additional information encoded in these maps, tracing both the interactions of CMB photons with matter along the line of sight \citep[e.g., CMB secondary anisotropies, see review by][]{aghanim08}, as well as the emission from objects between us and the surface of last scattering.  
The latter emission is dominated at short wavelengths by the cosmic infrared background (CIB; e.g., \citealt{dunkley11,thacker13,viero13a,mak17,reichardt20} and see \citealt{kashlinsky18} for a recent review)
and at longer wavelengths by synchrotron radiation, primarily from active galactic nuclei \citep[AGN; see e.g.,][]{dezotti10, everett20,gralla20}.  

In this work we focus on isolating signals from the CMB itself as well as two of the secondary CMB anisotropies: the kinematic and thermal  Sunyaev-Zel'dovich (kSZ and tSZ) effects. 
The kSZ and tSZ effects arise, respectively, from the first-order  \citep{sunyaev72,sunyaev80b} and second-order terms \citep{sunyaev70,sunyaev72} in the equations governing the scattering of CMB photons off moving electrons such as are found in galaxies, groups, and clusters  (see e.g., \citealt{birkinshaw99,carlstrom02} for a more detailed review). 
All three of these signals offer powerful probes of cosmology, with the tSZ and kSZ additionally probing a range of astrophysical processes. 

The CMB is, of course, one of the observational pillars of modern cosmology, with decades of observations having led to robust evidence for a spatially flat universe that arose from a hot Big Bang 
and whose composition and evolution today is governed primarily by dark energy and dark matter  \citep[e.g.,][]{mather94, hinshaw13, henning18,planck18-6,aiola20,adachi20}.
Observations of the SZ effects complement this picture by providing probes of the Universe at later times, particularly from the epoch of reionization to the present day. 
The tSZ effect, sourced primarily by the inverse Compton scattering of CMB photons off the hot gas in galaxy groups and clusters \citep{battaglia12,bhattacharya12}, enables the compilation of mass-limited cluster samples out to the epoch of cluster formation \citep[e.g., ][]{bleem15b,planck15-27,huang20,bleem20,hilton20}.  These samples, through growth of structure tests,  have been used to provide cosmological constraints competitive with those from  CMB and galaxy surveys  \citep{planck15-24,hasselfield13,bocquet19}. 
The tSZ power spectrum, bispectrum, and related observables also are highly sensitive to cosmology \citep[e.g.,][]{komatsu02,bhattacharya12,hill13,crawford14,george15,horowitz17,coulton18}. 
Finally the kSZ effect, while the faintest and thus far least characterized of these three observational probes, offers unique opportunities to explore the reionization history of the universe \citep{gruzinov98,shaw12,park13,battaglia13,smith17,reichardt20}  
including a new pathway to constraining the optical depth to reionization, $\tau$  \citep{ferraro18}. 
In combination with next-generation galaxy surveys---e.g., the Dark Energy Spectroscopic Instrument \citep{desi16}, Euclid \citep{amendola18}, SPHEREx \citep{dore14},  Vera Rubin Observatory's Legacy Survey of Space and Time \citep[LSST;][]{lsst09}, and the Roman Space Telescope \citep{spergel15}---it will also lead to new probes of gravity and dark energy  \citep{bhattacharya08,keisler13,mueller15a}, massive neutrinos \citep{mueller15b}, and other physics that influences the growth of structure \citep{bhattacharya07,sugiyama17,soergel18}. 

Observations of the SZ effects also can be used to constrain astrophysical processes. The redshift-independent nature of these signals makes them highly complementary to other physical probes both alone \citep{abazajian19}, and in combination with other observables \citep{siegel18,shitanishi18,okabe20,ruppin21}. 
In recent work, these signals have been used to provide insight into the ``missing baryon'' problem \citep[e.g.,][]{tanimura19,graaff19} by tracing the hot gas in filaments between massive galaxies, as well as to constrain astrophysical feedback and the thermal properties of both aggregate samples of galaxies \citep[e.g.,][]{soergel16,debernardis17,chavesmontero19,schaan20,amodeo20,vavagiakis21} and in high signal-to-noise measurements of individual clusters \citep{plagge10,planck12-5,romero17,ruppin18}.

Given the power of these sky signals to probe such a diverse and important range of scientific questions, there has been significant effort in developing techniques to isolate these signals into single-component maps \citep[see e.g.,][amongst many works]{remazeilles11,remazeilles11b,hurier13,bobin16,petroff20,abylkairov20}.   
In an idealized scenario, such maps would consist only of the desired observables of interest. 
However, the realities of noise, wavelength coverage (particularly restricted in ground-based observations by available atmospheric windows), limited knowledge of the spectral energy distributions of the various sky components \citep{tegmark98b} and/or incomplete differentiability between them, and the finite angular and spectral resolution of the available data all limit the fidelity of the signal map reconstructions. 
Such limitations typically necessitate tradeoffs between noise and signal purity in the reconstructed component maps. 

To optimize the use of available data, a number of works have focused on improving upon the component maps produced using data from \planck \  \citep[e.g., ][]{planck15-10,planck15-22,planck18-4,lenz19} by combining data from ground-based experiments with that from \planck \ to leverage the higher angular resolution of the terrestrial experiments and the superior noise performance and frequency coverage of \planck \  at large angular scales.  
An early example of this was \citet{crawford16}, in which data from \planck \ was combined with SPT observations to produce arcminute-scale emission maps of the Large and Small Magellanic Clouds for a range of emission spectra. 
\citet{aghanim19} combined  \planck \ data with 148 and 220 GHz ACT data from the 2008-2010 seasons \citep{swetz11,dunner12} to construct a $1\farcm{5}$ \ resolution tSZ component map using internal linear combination (ILC) techniques.
More recently,  \citet{madhavacheril20} used ILC techniques to construct $1\farcm{6}$ \ maps of both the tSZ and CMB/kSZ (including a variant of the latter from a ``constrained'' ILC where the tSZ was nulled) from \planck \ and  2100 square-degrees of  98 and 150 GHz data from the 2014-2015 ACTPol observing seasons, and  \citet{melin20} combined data from a public SPT-SZ release \citep{chown18} with that from \planck \ to produce a joint \planck +SPT cluster sample using matched filter techniques. 

In this work we describe the construction and public release of a series of maps constructed through the combination of data from the \planck \ mission with that from the 2500-square-degree SPT-SZ Survey \citep{story13}.  
These products include component maps of the tSZ and the CMB/kSZ useful for cluster detection and for cross-correlation analyses in which bias from foregrounds is expected to be minimal, as well as component maps where various contaminating sky signals have been significantly reduced or removed that are more optimal for e.g., CMB lensing reconstruction \citep{vanengelen14,osborne14,madhavacheril18} or other analyses in which the contaminating components could create significant biases in the quantities of interest.
We also provide with this release the 95, 150, and 220 GHz SPT-SZ maps used in this work.
Similar to the maps used in the production of the SPT-SZ cluster sample \citep{bleem15b}, these maps are both slightly deeper than and have higher resolution than the maps presented in \citet{chown18}, which conservatively used maps subject to the more stringent cuts from SPT power spectrum analyses \citep[e.g.,][]{george15} and were further degraded to $1\farcm{75}$ resolution.

This work is organized as follows. 
In Section \ref{sec:data} we describe the data used to construct and validate the component map products. 
In Section \ref{sec:ymap} we detail the process by which we construct the component maps and in Section \ref{sec:mapproducts} we describe the resulting products.  
In Section \ref{sec:char} we describe the results of several validation tests of the maps. 
Finally, in Section \ref{sec:summary} we summarize and conclude.  
All public data products in this paper are avaliable on the SPT page of the NASA/LAMBDA website\footnote{\webaddresslambda} and on the public SPT website.\footnote{\webaddress}

\section{Data and Processing} \label{sec:data}

The Compton-$y$ parameter and CMB/kSZ maps presented in this work are constructed using data from two surveys: the 2500-square-degree SPT-SZ survey and the \planck\ all-sky survey, as represented in the 2015 \planck\ data release.\footnote{In this work we make use of a number of temperature foreground products for which the \planck\ Legacy Archive \url{https://wiki.cosmos.esa.int/planck-legacy-archive/index.php/Foreground\_maps\#2018\_Astrophysical\_Components} recommends using products from the 2015 release. As such, we choose to make use of the 2015 maps in this work to ensure consistency in our analysis and note, given the updates detailed in \citet{planck18-3}, we expect minimal impact on analyses for which these maps will be most useful.}
The SPT and \planck \ data play highly complementary roles in this work: data from the SPT has higher resolution and lower noise at small scales, while the \planck \ data has lower noise at larger scales (multipole $\ell\lesssim 1700$) as well as enhanced frequency coverage. 
Additional data from the HI4PI survey \citep{hi4pi16} and the {\sl Herschel Space Observatory} SPIRE instrument \citep{pilbratt10,griffin03} were used to perform systematic checks of the map products. In this section we summarize these data and their processing as relevant to the component-map construction. 

\subsection{The South Pole Telescope SZ survey}\label{sec:survey}
The South Pole Telescope \citep{carlstrom11} is a 10 m telescope located approximately 1 km from the geographic South Pole at the National Science Foundation Amundsen-Scott South Pole station in Antarctica. To date, three different generations of instruments have been deployed on the telescope to conduct arcminute-scale observations of the millimeter-wave sky; the data used in this work were obtained between 2008 to 2011 with the SPT-SZ receiver.
This receiver was composed of six subarrays (or ``wedges") of feedhorn-coupled transition-edge bolometer sensors, with each wedge sensitive at either 95, 150, or 220 GHz. 
The resulting SPT-SZ survey covers a $\sim$2500 square-degree region extending from 20$^{h}$ to 7$^{h}$ in right ascension (R.A.) and $-65^{\circ}$ to $-40$$^\circ$ in declination (DEC); the full survey footprint was subdivided and observed in 19 separate fields ranging from $\sim$70 to 250 \sqdeg \ in size \citep{story13}. Data was acquired by scanning the telescope back and forth across each field in azimuth and then stepping in elevation, and then repeating the scan and step procedure until each field had been completely covered. One field, centered at R.A.=21h, DEC=$-$50$^{\circ}$, was observed in two different modes, with approximately 1/3 of the data obtained in the azimuth-scanning mode described above, while the remainder was obtained in an analogous process by scanning the telescope in elevation and then stepping in azimuth.  One complete pass of a field, lasting roughly two hours on average, is termed an \textit{observation}, and each field was observed at least 200 times. Two fields, the \texttt{ra23h30dec-55} and \texttt{ra5h30dec-55} fields, were observed for roughly twice the total time as the rest of the fields and, as a result, have roughly $\sqrt{2}$ lower noise.

\subsubsection{Map Making}\label{sec:mapmaking}

The general procedure for converting SPT observation data into field maps is presented in \citet{schaffer11}; we briefly summarize the process here.  
Maps are constructed from time-ordered bolometer data acquired during each scan across the field. Each detector's data is processed to remove sensitivity to the receiver's pulse tube cooler, cuts are applied based on both noise properties and responsiveness to sources, the data is rescaled based on the detector's response to an internal calibration source embedded in the telescope's secondary cryostat, and the data is filtered and binned into map pixels with inverse-variance weighting.

Filtering of time-ordered bolometer data consists of both an effective high-pass filter which  reduces sensitivity to atmospheric noise and a low-pass filter that prevents the aliasing of high-frequency noise when the detector data is binned into map pixels. 
For every scan across the field a low-order polynomial is fit and subtracted from each detector's time stream data, with the order
of the polynomial scaled to the length of the scan (roughly three modes per 10 degrees of scan for this work).
The effective $\ell$-space cutoff of this filter is roughly $\ell=50$ in the scan direction.
For the 220~GHz data, which is more sensitive to large-scale atmospheric fluctuations, we also fit and subtract
sines and cosines up to a frequency equivalent to $\ell = 200$ in the scan direction.
At each time sample in the observation, an isotropic common mode filter is applied to data from each wedge. 
At 95 GHz, this common-mode filter takes the form of subtracting the mean of all the detectors on the wedge from the time streams, while at 150 and 220 GHz, where the low-frequency atmospheric noise is more of an issue, two spatial gradients across the wedge are also subtracted from each detector's data. 
Finally, the data is low-pass filtered with a cutoff at angular multipole $\ell \sim$20,000.

In typical SPT analyses, to avoid introducing filtering artifacts around bright sources, sources brighter than a certain threshold are masked during filtering. 
In this work, because we combine SPT-SZ maps with \planck\ data on a Fourier-mode-by-Fourier-mode basis, 
we do not need to mask sources in the filtering steps that predominantly affect modes that can be filled in by \planck. 
This includes the common-mode filtering, which acts as an isotropic high-pass filter at the detector wedge scale (roughly half a 
degree). We do still mask sources in the polynomial subtraction (above a threshold of roughly 6 mJy at 150~GHz), because this acts as a high-pass in the scan direction only
and affects modes that oscillate slowly in the scan direction but quickly along the cross-scan direction and are thus not accessible
to the lower-resolution \planck\ observations. More generally, the fact that these modes are missing from both SPT-SZ 
and \planck\ data results in a small bias to the resulting Compton-$y$ and CMB/kSZ maps, which we discuss in Section~\ref{sec:trough}.

Following the filtering, the telescope pointing model is used to project inverse noise-variance weighted detector data into map pixels. 
For this work the SPT single frequency maps are made in the Sanson-Flamsteed projection \citep{calabretta02}, with a pixel scale of 0.25 arcmin. 
Individual observation maps are coadded with weights determined by the average detector noise performance in each observation. The coadded maps are absolutely calibrated using data from the \planck \ mission.  This calibration results in a $\sim$0.3\% uncertainty on the absolute calibration of the SPT data; see \citet{hou18} for details on the calibration procedure.
To avoid boundary effects in the map combination process (Section \ref{sec:ymap}) we ``pad'' the fields by 25\% in area by coadding with their neighbors and require coverage at all 3 SPT-SZ frequencies.\footnote{Owing to asymmetry in wedge locations in the SPT-SZ focal plane (Section \ref{sec:survey}), the 95 and 220~GHz data have slightly reduced coverage at opposite field edges.} These requirements reduce the component map areas  compared to e.g., the area searched for clusters in \citet{bleem15b} by slightly shrinking the exterior boundary of the survey; the total area is reduced by $\sim$3\%.

In this work we use the same data cuts as in \citet{bleem15b}, leading to depths in the final coadded field maps at  $4000 <\ell< 5000$ of approximately  37, 16, and 65 \muk-arcmin at 95, 150, and 220 GHz respectively, using the Gaussian beam approximation estimation of \citet{schaffer11}. For a complete listing of field depths, see e.g., Table 1 in \citet{bleem15b}.

\subsubsection{Noise PSDs}\label{sec:noisepsds}
As in a number of previous SPT publications \citep[e.g.,][]{staniszewski09}, we use a resampling technique to estimate two-dimensional map noise power spectral densities (PSDs). In each resampling, individual observations of a given SPT-SZ field are randomly assigned a sign of $\pm$1 and coadded; sky signal is highly suppressed in the resulting coadded maps, and we approximate them as pure noise. We compute the PSD of each signal-free coadd, and we use the average of these jackknife PSDs to provide an estimate of instrumental and residual atmospheric noise in the filtered SPT-SZ maps.

\subsubsection{Angular Response Function}\label{sec:beams}
The SPT-SZ sky maps created as described above are biased representations of the true sky, as the data has been modified by both the finite resolution of the instrument and the filtering processes described above. At each frequency, the instrumental response function (i.e., beam) is to first order well represented as an azimuthally symmetric Gaussian with full width at half maximum (FWHM) equal to $1\farcm{6}$, $1\farcm{1}$, and $1\farcm{0}$ \ at 95, 150, and 220 GHz respectively. As described in \citet{keisler11} and \citet{schaffer11}, the beam of the SPT is characterized using a combination of emissive sources in the survey fields and dedicated planet observations. In the process of creating the component maps (described in Section~\ref{sec:ymap}), we use the azimuthally averaged Fourier space beams $B(\ell)$.

To characterize the effect of our map filtering we simulate the filter response function. Following \citet{crawford16}, 100 simulated skies are constructed from white noise convolved with a $0\farcm{75}$ \ Gaussian beam. Each sky realization is ``mock observed" to obtain detector timestream data using the telescope pointing model. These time-ordered data are cut, weighted, and filtered in an identical process to the real data to form individual-observation maps. These individual-observation maps are combined with same weighting as the real data to make final coadded mock-observed maps. For each sky realization the estimate of the filter response function is then determined as the ratio of the two-dimensional power spectrum of the coadded map to the known input power spectrum. The 100 independent ratios are averaged and this averaged ratio is used as the estimate of the 2D Fourier-space filter response function. Because the filter transfer function is highly similar among all fields except the ``el-scan'' field (see Section~\ref{sec:survey}), we ran the mock observations only on the el-scan field and a single standard-observation field; for the other 17 standard-observation fields we used a regridded version of the standard-field filter transfer function.

 \subsubsection{Bandpasses}\label{sec:bandpasses}
The process of separating a component map signal from other sky signals and noise relies on a priori knowledge of the desired signal 
spectrum and, by extension, knowledge of the spectral response of the instruments used to collect the data.
The bandpasses of the SPT-SZ detectors were characterized using a Fourier transform spectrometer. Spectra were measured for $\sim$50\% of all bolometers on each wedge and, as the spectra were found to be highly uniform (the bandpasses of the SPT-SZ detectors are set by a low-pass metal-mesh filter covering each wedge and a high-pass filter from the waveguide cutoff frequency of precision machined feedhorns), a weighted-average spectrum for each frequency was computed.  Weights were set by the inverse square of each detector's noise-equivalent temperature. 

As the SPT-SZ bands are not infinitely narrow, the response to signals with different spectral energy distributions 
cannot be characterized by a single center frequency. We follow, e.g., \citet{reichardt20} and calculate an effective band 
center for every sky component in the signal model (see Section~\ref{sec:ymap}). We define the effective band center 
for a given source spectrum as the frequency at which 
the conversion between CMB fluctuation temperature and intensity is equal to the value integrated over the true SPT-SZ bands,
weighted by the source spectrum.
For a thermal SZ spectrum this leads to effective band centers of approximately 99.6, 155.5, and 220.0 GHz for the three SPT-SZ frequencies. 

Another consequence of the SPT (and \planck) bands not being infinitely narrow is that the angular response function may not be constant
over the band, resulting in effective beams that are different for different spectral energy distributions (SEDs). We have calculated 
the maximum bias expected for any of our output component maps from this effect at any value of multipole $\ell$ and find it to be 1.2\%
(with the typical value over the full $\ell$ range much lower), and as a 
consequence we ignore this effect in our analysis.
 
\subsection{\planck }\label{sec:planck}

The European Space Agency's \planck \ satellite was launched in 2009 with the primary mission of conducting sensitive, high-resolution observations of the CMB \citep{tauber10,planck11-1}.
Full sky maps from 25-1000 GHz were produced over the course of its multi-year mission. 
We make use of products from the 885-day High Frequency Instrument (HFI) cold mission that are included in the \planck \ 2015 public release \citep{planck15-1}. HFI data in the 100, 143, 217, and 353 GHz channels are used in the map making process, and data from the 545 GHz channel is used for several systematic checks. 
Five surveys, each covering a large fraction of the sky, were conducted over the course of the cold HFI mission; both full and half-mission maps produced from these surveys are used in this work. 
 
\subsubsection{Maps, Angular Response Function, Noise Treatment}\label{sec:planckchar}

For each channel the time-ordered data is converted into maps as described in \citet{planck13-8} and updated in \cite{planck15-8}. As detailed in the latter work, the data are calibrated at 100-353 GHz using the time-variable CMB dipole and at 545 GHz using planetary emission from Uranus and Neptune (see also further discussion of the calibration using planets in \citealt{planck17-52}). The absolute calibration of the full mission HFI maps is determined to [0.09, 0.07, 0.16, 0.78, 1.1(+5 model uncertainty)] \% accuracy at [100, 143, 217, 353, 545] GHz.

The \planck \ maps are processed such that the transfer function can be modeled as an ``effective" beam---accounting for the telescope optics, survey scan strategy, and data processing---convolved with the map pixelization window function \citep{planck13-7}. We use the azimuthally symmetric  ``Reduced Instrument Model" (RIMO) effective beam window functions derived from the 75\% of the sky outside the Galactic plane.\footnote{\url{https://wiki.cosmos.esa.int/planckpla2015/index.php/Effective_Beams}} We also make use of the beam files provided with \citet{planck20-57} to extend the 217 and 353~GHz beam from $\ell_\textrm{max}=4000$ to $\ell_\textrm{max}=8192$. 
Using a Gaussian approximation the typical FWHM of these effective beams is [9.69, 7.30, 5.02, 4.94, 4.83] arcminutes at [100, 143, 217, 353, 545] GHz \citep{planck15-7}.
These maps are provided in HEALPix\footnote{http://healpix.sourceforge.net} format \citep{gorski05} with $N_{\mathrm{side}}=2048$, corresponding to $1\farcm{7}$ \ pixels.  The RIMO also provides the \planck \ bandpasses \citep{planck13-9}. 

While the time-ordered \planck \ data is approximately ``white'' with a ``$1/f$" contribution at lower frequencies \citep{planck11-4}, data processing and map projection can introduce correlations between map pixels \citep{planck15-8}. 
In this work, as in \citet{crawford16}, we ignore these non-idealities, and model the \planck \ noise as white noise with levels set on a field-by-field basis. 
This treatment will not bias the resulting component maps and should have only a minor effect on the optimality of the band combination, as the \planck \ data primarily serves to fill in the large angular scale modes that have been removed or down-weighted by the SPT filtering. 
Noise levels are estimated using the square root of the mean of intensity-intensity variance maps in each SPT field. 
These noise levels vary typically vary by 5--10\% rms across the SPT-SZ fields (but up to 30\% near the South Ecliptic Pole, where
the \planck\ observation strategy results in very deep coverage). The mean \planck\ noise level varies by up to a factor of two 
between SPT-SZ fields (again particularly near the Ecliptic Pole), but this is taken into account as both the CMB/kSZ and Compton-$y$ maps are constructed independently in the 19 SPT-SZ
fields.

The \planck \ and SPT data are combined on a field-by-field basis, and then the maps constructed from individual fields are stitched together to form full SPT-SZ survey CMB/kSZ and Compton-$y$ maps. This field-by-field procedure enables us to more optimally combine the \planck \  and SPT datasets given noise variations across each survey. 
To match the SPT map projections we first rotate the \planck \ maps from the Galactic to Celestial coordinate system using the HEALPix \texttt{rotate\_alm} function. The \planck \ maps are then projected using the Sanson-Flamsteed projection to match each SPT field's pixelization. 

\subsection{Treatment of Bright Sources in SPT and \planck \ Data}\label{sec:inpaint}
While data from \planck \ is used to fill in missing modes at low $\ell$ removed during filtering of the SPT maps (see e.g., Sections \ref{sec:mapmaking}, \ref{sec:trough}) this compensation is incomplete for time-variable signals or when a foreground source has 
an SED that is 
very different from the SED of the desired signal component.
The former effect is particularly important for bright sources (dominated by flat- or falling-spectrum blazars at the SPT wavelengths), whose flux could vary significantly between the time of the \planck\ and SPT observations. 
Both of these effects can cause a mismatch in the power in the SPT and \planck\ maps, resulting in artifacts in the output map.
To mitigate such artifacts we ``paint in'' the regions around the brightest sources in the maps.  
For sources detected at $>250$ mJy at 150~GHz in the SPT maps (as well as 6 additional sources from the \planck \ 143~GHz source catalog \citep{planck15-26} above this threshold in the \planck \ but not SPT data) we fill in regions of radius 20\arcmin \ around each source location in all of the maps  using the mean value of the sky temperature computed from annuli extending 15\arcmin \ from the inpainting radius.
We do the same for sources above 150 mJy at 150/143~GHz but with a 10\arcmin \ radius.
These painted regions are masked in the provided source mask.

\subsection{Herschel}\label{sec:herschel}

We tune the Compton-$y$ reconstruction parameters to minimize the correlation with the CIB by using data from the \textit{Herschel Space Observatory} \citep{pilbratt10} SPIRE instrument \citep{griffin03}.
The \textit{Herschel} data consists of observations at 500, 300, and 250 \um that were acquired over a $\sim$90 square-degree patch centered at (RA,DEC)=(23h30m,~$-55^{\circ}$) in the SPT-SZ survey  under an Open Time program (PI:~Carlstrom). In this study we make use of the 500 \um \ maps. The \textit{Herschel}/SPIRE angular resolution is superior to that of both SPT and \planck---the effective resolution of the 500 \um maps is $36\farcs{6}$ ---enabling us to test CIB contamination at small scales. Further details on these maps and their processing can be found in \citet{holder13}.  

\subsection{HI4PI}
To validate the removal of Galactic dust from the maps at large scales we make use of data from the HI4PI survey of neutral atomic hydrogren \citep{hi4pi16}. Such HI survey data has long been known to be well-correlated with infrared-to-millimeter-wavelength dust emission \citep[e.g.,][]{boulanger88,boulanger96,lagache03,planck13-7}.
The survey maps in the SPT region are from the third revision of the Galactic All Sky Survey \citep[GASS,][]{mcclure-griffiths09,kalberla10,kalberla15} which was conducted in 2005-2006 with the Parkes Radio Telescope. These maps have an angular resolution of FWHM=16$\farcm{2}$ and an rms noise temperature of approximately 43~mK. 
We utilize the N$_{\rm HI}$ column density maps provided in HEALPix format. 
These were constructed via integrating over the full velocity range of GASS (with absolute radial velocity $\leq$470 km s$^{-1}$).  
Further details on these data products can be found in \citet{hi4pi16}.

\section{Construction of the CMB/kSZ and Compton-$y$ Component Maps}\label{sec:ymap}
In this work we assume the \planck \ and SPT temperature maps to be composed of 
sky signal contributions from primary CMB temperature fluctuations, tSZ, kSZ, 
radio galaxies, 
the dusty galaxies that make up the CIB, 
and Galactic dust; and noise contributions from the instrument and---for SPT---the atmosphere.
In this section we describe the procedure for extracting an unbiased component map 
from individual frequency maps; the models we adopt for the instrumental, atmospheric, and astrophysical
contributions to the variance in the individual maps and covariance between them; 
the treatment of the ``missing modes'' in output component maps; 
and our procedure for suppressing CIB contamination in the ``minimum-variance'' Compton-$y$ maps.

\subsection{Linear Combination Algorithm}
\label{sec:lincomb}
We use a linear combination of individual frequency maps to construct the Compton-$y$ map. 
The practice of linearly combining mm-wave/microwave maps to extract individual sky components
is common in the CMB field, and there are many approaches (see, e.g., the review by \citealt{delabrouille09b}).
All mm-wave/microwave linear-combination (LC) algorithms have a common goal, 
namely to combine a set of single-frequency maps to produce a map with unbiased response to a 
signal with a known frequency spectrum. LC algorithms also generally seek to minimize the variance
in the output map from instrument noise, atmospheric fluctuations (if the data is from a ground-based 
experiment), and sources of astrophysical signal other than the target signal. For some applications, 
it is preferable to minimize the total variance in the output map, while for other applications it is preferable
to explicitly null the response of the output map to one or more signals (assuming the SEDs
of those signals are known perfectly). An example of the latter type of map is a map of the CMB temperature
fluctuations that has null response to tSZ, which can be a significant contaminant to measurements of CMB  
lensing and the kSZ effect \citep[e.g.,][]{madhavacheril18,baxter19,raghunathan19a}.

Another choice that must be made in constructing the LC is how to characterize the variance in the
individual frequency maps and the covariance between them. One choice is to use models of the 
various sources of power in the maps, including instrumental and atmospheric noise as well as astrophysical
signals. The downside of this approach
is that if the models are not perfect, there will be excess residual variance in the output component map(s). 
Another approach, referred to as ``internal linear combination" or ``ILC'' in the literature, estimates 
the variance and covariance from the individual frequency maps themselves. The downside to 
this approach is that it results in a biased output map \citep[e.g.,][]{delabrouille09}. This ``ILC bias'' can be mitigated by averaging 
the covariance estimates over many real-space or Fourier-space pixels. The ILC approach with 
Fourier-space averaging was used to
produce the ACTPol+\planck\ Compton-$y$ map in \citet{madhavacheril20}.
We choose to adopt the first 
approach, preferring to incur a potential noise penalty than a bias in the output map. 
We estimate the potential excess variance from imperfect modeling in Section~\ref{sec:residual_contamination}.

Operationally, this approach is
identical to the map combination procedure used in the construction of the SPT-SZ cluster sample 
(e.g., \citealt{melin06,bleem15b}), replacing the  $\beta$-profile filter used to optimize cluster identification 
with a profile with a flat response as a function of $\ell$. 
In this approach, each single-frequency map at sky location $\mathbf{n}$ is characterized as a sum over the various sky signal components, convolved with the appropriate beam and transfer function (see Section \ref{sec:beams}), plus instrumental---and in the case of SPT, atmospheric---noise.
\begin{equation}
T(\nu_i, \mathbf{n}) =  R(\nu_i, \mathbf{n}) * \sum_j f_j(\nu_i) S_{j}(\mathbf{n}) + n_\mathrm{noise}(\nu_i, \mathbf{n}),
\label{eqn:maps}
\end{equation}
where $R(\nu_i, \mathbf{n}) = B(\nu_i, \mathbf{n}) * F(\nu_i, \mathbf{n})$ is the convolution of the real-space 
beam and filter kernel, and $f_j(\nu)$ is the SED of signal $S_j$.
We choose to work in two-dimensional Fourier space, 
using $\mathbf{l}$ as the wavenumber-like Fourier conjugate of sky location $\mathbf{n}$, in which 
case we can write Equation~\ref{eqn:maps} as 
\begin{equation}
T(\nu_i, \mathbf{l}) =  R(\nu_i, \mathbf{l}) \ \sum_j f_j(\nu_i) S_{j}(\mathbf{l}) + n_\mathrm{noise}(\nu_i, \mathbf{l}),
\label{eqn:mapsf}
\end{equation}
where the $R(\nu_i, \mathbf{l}) = B(\nu_i, \ell \equiv |\mathbf{l}|) F(\nu_i, \mathbf{l})$ is the product of the Fourier-space 
beam window function (which we approximate as azimuthally symmetric) and filter transfer function
(which is manifestly anisotropic for SPT-SZ).

As shown in \citet{melin06} and (in an alternate derivation) in Appendix~\ref{appendix:derivation}, under the 
assumption that the instrumental/atmospheric and astrophysical noise models are correct, the combination of 
the data that yields the minimum-variance, unbiased map of signal $S$ with SED $f(\nu)$ in two-dimensional Fourier space is:
\begin{equation}
\bar{S}(\mathbf{l}) = \sum_i \psi(\nu_i, \mathbf{l}) \; T(\nu_i, \mathbf{l}),
\label{eqn:ymap}
\end{equation}
where 
\begin{equation}
  \psi(\nu_i, \mathbf{l}) = \sigma_{\psi}^{2} (\mathbf{l}) \sum_j \mathbf{N}_{ij}^{-1}(\mathbf{l}) f(\nu_j) R(\nu_j, \mathbf{l}),
\label{eqn:psi}
\end{equation}
is the weight contributed to the output map $\bar{S}$ by band $\nu_i$ at angular frequency $\mathbf{l}$, 
$\sigma_{\psi}^2 (\mathbf{l})$ is the predicted variance of the output map at angular frequency $\mathbf{l}$, given by
\begin{equation}
  \sigma_{\psi}^{-2} (\mathbf{l}) = \sum_{i,j} f(\nu_i)R(\nu_i, \mathbf{l}) \mathbf{N}_{ij}^{-1}(\mathbf{l}) f(\nu_j) R(\nu_j, \mathbf{l}),
\label{eqn:sigmapsi}
\end{equation}
and $\mathbf{N}_{ij}(\mathbf{l})$ is the two-dimensional Fourier space version of the band-band, pixel-pixel covariance matrix,
including contributions from instrumental and atmospheric noise, and 
from sky signals other than the signal of interest. The model used to 
construct $\mathbf{N}$ is presented in Section~\ref{sec:skymodel}.
We normalize $\mathbf{N}$ so that the squared, 
azimuthally averaged Fourier-domain power as a function of $\ell = | \mathbf{l}|$ is equivalent to $C(\ell)$ in the flat-sky limit.

To make a map of signal $S$ with some other signal formally nulled, we simply replace $f$ in the previous equations
with an $N_\mathrm{bands} \times N_\mathrm{components}$ matrix $f_i(\nu_j)$ encoding the spectral behavior 
of all signals for which we desire independent output maps. The band-weighting function $\psi$
then becomes an $N_\mathrm{bands} \times N_\mathrm{components}$ matrix at each two-dimensional Fourier location, 
and the output component maps $\bar{S}_i$ are given by
\begin{equation}
\bar{S}_i(\mathbf{l}) = \sum_j \psi_i(\nu_j, \mathbf{l}) \; T(\nu_j, \mathbf{l}),
\end{equation}
where 
\begin{equation}
  \psi_i(\nu_j, \mathbf{l}) = \sum_{k,m} C_{\psi,ik} (\mathbf{l}) \mathbf{N}_{km}^{-1}(\mathbf{l}) f_j(\nu_m) R(\mathbf{l}, \nu_m),
\label{eqn:psimult}
\end{equation}
and
\begin{equation}
  C^{-1}_{\psi,ij} (\mathbf{l}) = \sum_{k,m} f_i(\nu_k)R(\nu_k, \mathbf{l}) \mathbf{N}_{km}^{-1}(\mathbf{l}) f_j(\nu_m) R(\mathbf{l}, \nu_m).
\end{equation}
We show in Appendix~\ref{appendix:equiv} that for the particular case of two components this band-weighting function is
identical to Equation 3 in \citet{madhavacheril20}, as derived in, e.g., \citet{remazeilles11}.

There is an important assumption implicit in this formulation, namely that the statistical behavior of the instrumental/atmospheric 
noise and the astrophysical contaminants does not vary across the maps. This allows the noise and contaminants to be 
meaningfully represented in Fourier space and (under the assumption of Gaussianity) means that the Fourier components
at different frequencies are independent, so that $\psi$ can be computed independently at every value of $\mathbf{l}$.
For many existing and upcoming datasets, particularly those covering a large fraction of the sky, these conditions
are not met, and a spatially varying approach such as needlet ILC \citep{delabrouille09} must be employed.
This is, however, a good approximation for the noise and foregrounds in this analysis. 
The noise in the SPT-SZ survey is well approximated as statistically uniform, particularly because we perform the component separation
in each of the 19 SPT-SZ fields individually. The SPT-SZ noise in these fields only varies slightly as a function of 
declination, and the \planck\ noise across any one of these $\sim$200 deg$^2$ regions is fairly uniform as well
(see Section~\ref{sec:planckchar}). 

Most of the relevant 
astrophysical components are both statistically isotropic across the SPT-SZ fields and Gaussian-distributed, with the 
notable exceptions of Galactic dust and synchrotron and individual bright emissive sources and galaxy clusters. The SPT-SZ fields are at 
high Galactic latitude, and at the angular scales relevant for tSZ studies Galactic foregrounds are negligible in these 
fields---though at the very largest angular scales, Galactic dust is clearly visible in the resulting Compton-$y$ map if we do not correct for it, 
see Section~\ref{sec:dust} for details. Meanwhile, we mask all emissive sources down to flux densities of $\sim$6 mJy at
150 GHz, significantly reducing the statistical anisotropy and non-Gaussianity of that component. We note that the 
contribution of the tSZ to the covariance is not actually relevant to this work, because in all of our output maps the tSZ is
either the signal of interest or one of the signals being nulled.

\subsection{Sky and Noise Model}
\label{sec:skymodel}
In this section, we describe the model used to construct the 
spectral behavior matrix $f_i(\nu_j)$ and the 
two-dimensional Fourier-space covariance matrix $\mathbf{N}_{ij}(\mathbf{l})$  
used in the LC algorithm described in Section~\ref{sec:lincomb}.
$\mathbf{N}_{ij}(\mathbf{l})$ includes contributions from astrophysical sources and instrumental and atmospheric noise,
and it includes covariance between maps at different observing frequencies.
We model the amplitude and angular power spectra of the astrophysical contributions to the SPT-SZ and \planck\ maps based on recent power spectrum results from the SPT and \planck \ collaborations, while the noise model is based on the estimates
described in Sections~\ref{sec:noisepsds} and \ref{sec:planckchar}.

\begin{figure*}
\begin{centering}
\includegraphics[width = 0.9 \textwidth]{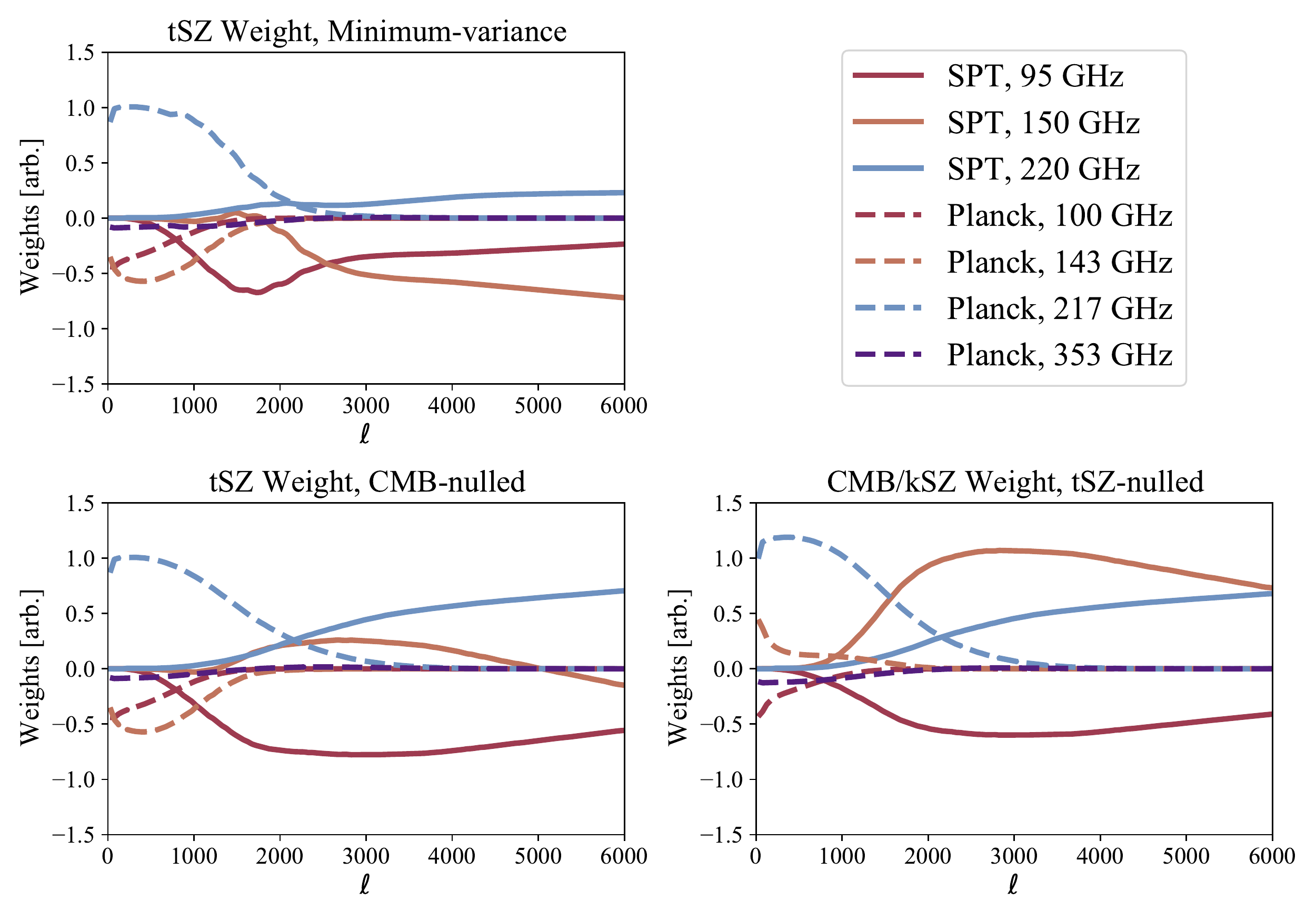} 
\caption{
Relative azimuthally averaged band weights for the minimum-variance thermal SZ Compton-$y$ map (\textit{upper left}), 
the CMB-nulled $y$ map (\textit{lower left}), 
and the tSZ-nulled CMB/kSZ map (\textit{lower right}).
}
\label{fig:weights}
\end{centering}
\end{figure*}

The contributions to the sky signal are modeled as follows: 
\begin{itemize}
\item 
The tSZ
appears as a spectral distortion of the CMB spectrum in the direction of energetic electrons, such as those found in galaxy groups and clusters.  In this work we follow  \citet{sunyaev72,sunyaev80} where this distortion at a given frequency, $\nu$, is given by: 
\begin{equation}\label{eq:sze}
\begin{split}
\Delta T(\nu )&= \Tcmb \ \fsz(x)\int n_\mathrm{e} \frac{k_\mathrm{B}T_\mathrm{e} }{m_\mathrm{e}c^{2}} \sigma_\mathrm{T} dl  \\
  &\equiv  \Tcmb \ \fsz(x) \ \ysz,
\end{split}
\end{equation}
where the integral is along the line of sight,   $\Tcmb=2.7260\pm 0.0013$~K is the mean CMB temperature \citep{fixsen09b}, \mbox{$x \equiv h\nu$/$k_\mathrm{B} \Tcmb$}, $k_\mathrm{B}$ is the Boltzmann constant, $c$ the speed of light, $n_\mathrm{e}$ the electron density, $T_\mathrm{e}$ the electron temperature, $\sigma_\mathrm{T}$ the Thomson cross-section,  and
 $\fsz(x)$ is the frequency dependence of the effect relative to
the spectrum of fluctuations of a 2.73K blackbody (note that we assume that all input maps are calibrated 
such that they have unit response to CMB fluctuations):  
\begin{equation}
\fsz(x) = \left ( x\frac{e^{x} + 1}{e^{x}-1} - 4 \right ) (1 + \delta_\mathrm{rc}), 
\end{equation} 
with $\delta_\mathrm{rc}$ encompassing relativistic corrections \citep[e.g.,][]{wright79,nozawa00,chluba12}.
While these relativistic corrections can have a significant impact on the tSZ signal at e.g., the locations of hot clusters (e.g., \citealt{hurier16}), in this work, at small angular scales where this new Compton-$y$ map will offer significant advantages over the \planck \ data previously available in this region, omitting these corrections has negligible impact on the weights in the band combination.   
At high $\ell$ (see Figure \ref{fig:weights}) the $y$ map construction is dominated by a combination of the SPT 95 and 150 GHz channels and the difference in the relative strength of the tSZ signal in these bands when including relativistic  corrections is small (for example the ratio of $\fsz$ in the 95/150 GHz bands for a 15 keV cluster is 1.70 when including relativistic corrections compared to 1.71 without). 
The amplitude at the location of hot gas will be biased low however, e.g., 8\% (3.5\%) for  15 (5) keV gas \citep{nozawa00}.

\item The primary CMB is modeled following the power spectrum results presented in \citet{planck18-5}. In particular we 
use the temperature power spectrum predicted by the best-fit cosmological model to the \planck\
\textsc{plikHM\_TTTEEE\_lowl\_lowE\_lensing} data set.
\item The CIB is modeled based on the results of \citet{reichardt20} from the SPT-SZ and SPTpol surveys. 
It is parameterized into two components: a Poisson component whose power spectrum is independent of angular scale and a clustered component that follows the one and two halo clustering template of \citet{viero13a}.
Both components are assumed to have a scaling with frequency that follows a modified black body: 
\begin{equation}
\eta_{\nu}  = \nu^\beta B_\nu(T)
\end{equation}
where $B_\nu$ is the black-body spectrum for a temperature $T$  and $\beta$ is the dust emissivity index.  For a fixed dust temperature of 25 K, \citet{reichardt20} constrains $\beta_{\rm P}= 1.48 \pm 0.13$ for the Poisson component and $\beta_{\rm cl} = 2.23 \pm 0.18$ for the clustered term
with 150~GHz power at $\ell=3000$ of $D^{\rm P}_{3000}=7.24 \pm 0.63$~$\mu$K$^{2}$ for the Poisson component and $D^{\rm one-halo}_{3000}=2.21 \pm 0.88$~$\mu$K$^{2}$ and $D^{\rm two-halo}_{3000}=1.82 \pm 0.31$~$\mu$K$^{2}$ for the one- and two-halo terms, respectively.\footnote{Where $D_{3000}$ refers to power at $\ell=3000$ in the convention $D_\ell=\frac{\ell(\ell+1)}{2\pi}C_\ell$.}

\item We adopt a two-fold approach to reduce contamination from Galactic cirrus.  As described below in Section \ref{sec:dust}, at large angular scales we subtract off a template of the dust emission from each of the single-frequency maps used in the analysis.  At smaller angular scales we incorporate the dust signal as a component in our noise covariance matrix. We assume that this cirrus signal is well represented by a modified blackbody that is 100\% correlated between bands with a dust temperature of 19 K \citep{viero19}, a dust emissivity of $\beta_\mathrm{cirrus} = 1.89$ \citep{martin10}, and a spatial dependence of $D_\ell \propto \ell^{-1.2}$ which we further reduce at large angular scales to correct for the power removed by the template subtraction. The amplitudes of the cirrus signal at each frequency is set by scaling the results of \citet{george15}. 
\item We include a Poisson component for the radio source power based on the results of \citet{reichardt20} which found a spectral index,  $\nu^{\alpha}$, $\alpha=-0.76$ and amplitude $D_{3000}=1.01 \pm 0.17$~$\mu $K$^{2}$  for sources fainter than 6.4 mJy at 150 GHz in SPT data. 
We also provide a mask with the location of sources detected above this threshold \citep{everett20}.  
After applying this mask we discuss residual contamination from sources below this threshold in Section \ref{sec:residual_contamination}.
\end{itemize}

As discussed in the previous section, 
the intrinsic two-dimensional fluctuation power from all of these signals is expected to be statistically isotropic
except for Galactic foregrounds and individual bright emissive sources, which we argue contribute negligibly 
to the covariance matrix. Sky signals that contribute significantly to the covariance are also expected to be
isotropic on the sky (no preferred direction). In this case, we can write
\begin{equation}
\langle S_{j}(\nu_i, \mathbf{l_1}) \ S_{j}(\nu_k, \mathbf{l_2}) \rangle = C(\nu_i, \nu_j, \ell_1 = |\mathbf{l_1}|) \ \delta(\mathbf{l}_2=\mathbf{l}_1)
\end{equation}
and characterize the signal power with an independent band-band covariance matrix only dependent on $\ell \equiv |\mathbf{l}|$.
As discussed in Sections~\ref{sec:noisepsds} and \ref{sec:planckchar}, we can also approximate the SPT-SZ and \planck\
noise as statistically isotropic (and thus uncorrelated between different $\mathbf{l}$ modes).
Because of the combination of the SPT scan pattern and low-frequency noise from atmosphere and other sources, 
the noise part of the covariance matrix is not isotropic on the sky (modes that oscillate quickly along the SPT scan 
direction are less noisy than those that oscillate slowly along that direction). For this reason we keep the full two-dimensional
noise covariance $ \mathbf{N}_{ij}^\mathrm{noise} (\mathbf{l})$.

\subsection{Galactic Dust}\label{sec:dust}
While the SPT-SZ fields were generally chosen to target areas of high Galactic latitude, there is still appreciable emission at large angular scales from Galactic dust.  As this emission is spatially variable across the survey footprint we choose to subtract a template of this contaminant from each frequency map before including it in our Fourier-based map construction.  We explored constructing this template using data from the HI4PI survey \citep{hi4pi16}, the   \planck \  thermal dust map constructed using the generalized needlet internal linear combination (GNILC) technique \citep{planck16-47}, and the CMB-subtracted \planck \  545 GHz ``foreground'' map constructed using the COMMANDER algorithm \citep{planck15-10}\footnote{See also details at \url{https://wiki.cosmos.esa.int/planckpla2015/index.php/CMB_and_astrophysical_component_maps}}.

Ultimately we settled on using the 545 GHz foreground map smoothed to a FWHM  of 16\arcmin \  to construct our template, validating the dust removal using the HI4PI survey. We note that while the GNILC thermal dust map has superior dust modeling, its resolution is spatially varying (with FWHM ranging from  5\arcmin \ to $21\farcm{8}$), and---as the dust levels in the SPT footprint are generally low and well-subtracted with our uniform resolution template---the results are changed negligibly in adopting the smoothed 545 GHz map, the uniform resolution of which simplifies modeling.    There is a small amount of tSZ in the 545 GHz COMMANDER map, both from the true tSZ signal at 545 GHz and from tSZ leaked from other frequencies in the process of removing the CMB signal using multifrequency \planck \ data \cite[see e.g., discussion in ][]{chen18}. We use simulations to quantify the level of tSZ in the 545 GHz COMMANDER map and the degree to which this biases our final recovered Compton-y map. We find that the bias to the recovered Compton-y signal is $\sim1\%$ at $\ell<200$ and much lower at higher $\ell$. In practice, this boost partially offsets the small loss in power caused by our low-$\ell$, high-$k_x$ filtering (see Sections \ref{sec:trough},\ref{sec:troughbias}). 

To determine the scaling between the dust template and each \planck \ frequency map we first project both the  545 GHz foreground template map and the COMMANDER foreground map for the \planck \ frequency of interest onto each of the SPT fields. We smooth each map to a resolution of 16\arcmin \ as we are interested in determining the scaling between Galactic dust and the \planck \ frequency maps and do not want to include the smaller-scale CIB in our fits and subtraction here. We next rebin the data to 4\arcmin\ pixels and mask bright emissive sources detected at $>150\sigma$ in SPT data at 95 or 150 GHz with degree-radius masks to reduce spurious correlations between the maps. We conduct an outlier-resistant polynomial fit over all the SPT-SZ fields to find the global relation (one per frequency for the whole survey) between the intensity in the template and the various smoothed foreground maps, finding a cubic polynomial to provide the best fit.  The smoothed 545 GHz foreground maps are then scaled by this relation and subtracted off from each of the \planck \ full sky maps before they are used in the LC described above.

We also fit for a similar correlation using the SPT data, this time additionally filtering the 545~GHz template map with the SPT transfer function. However, the high-pass filter applied to the SPT maps has already effectively removed the majority of the large scale emission and no significant correlation was detected, even reducing the smoothing of the maps to a FWHM of 10\arcmin. Given the small SPT data contribution at the $\ell$ ranges affected (see e.g., Figure \ref{fig:weights}), we choose to simply ignore the low-$\ell$ dust contribution to the SPT maps (but do include a dust covariance matrix as a source of noise in the map combination, as detailed above).  

\begin{figure}[t]
\begin{centering}
\includegraphics[width=\columnwidth]{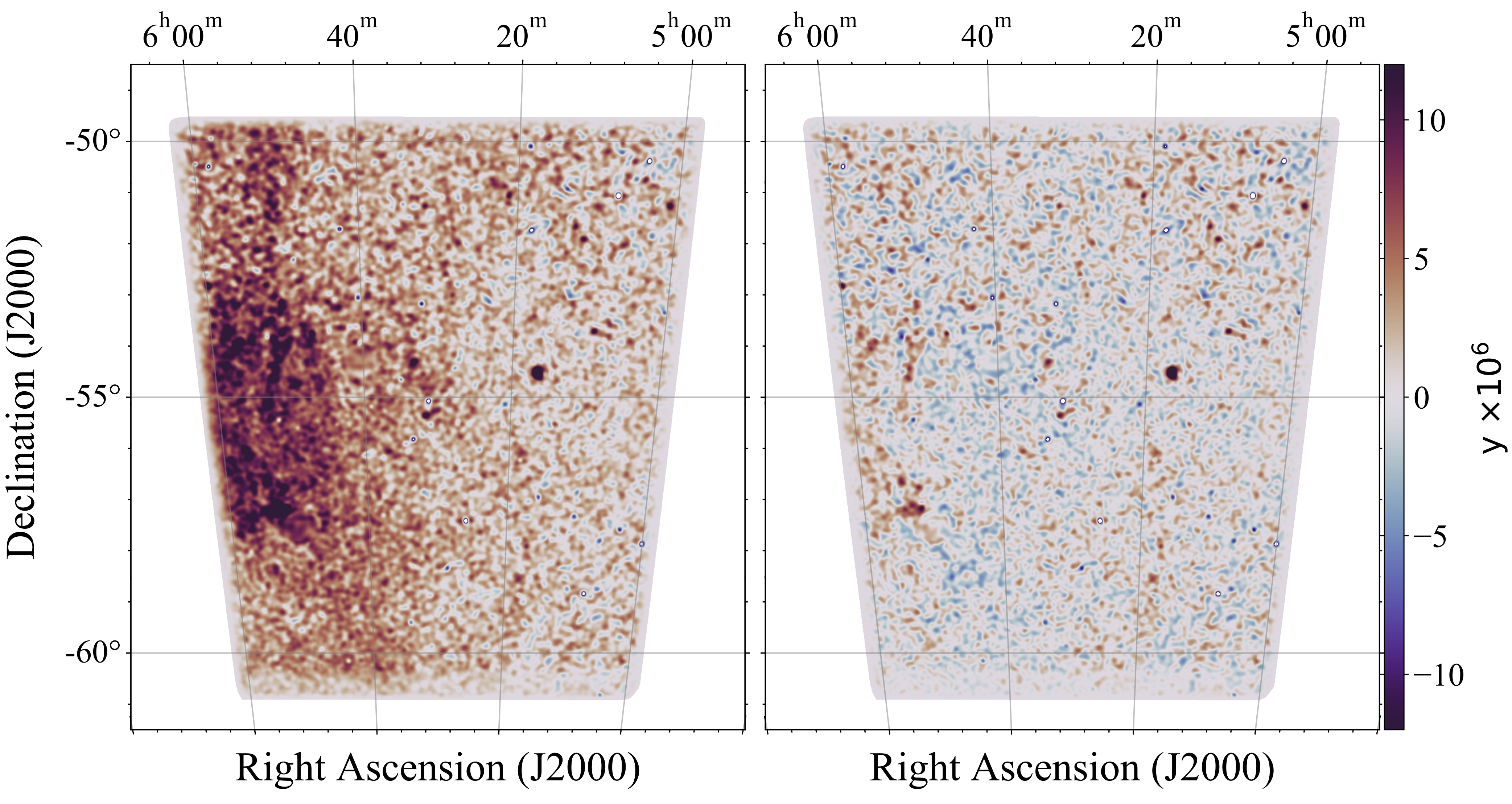} 
\caption{
\textit{Left panel:} 
Minimum-variance Compton-$y$ map for the SPT-SZ \texttt{ra5h30dec-55} field constructed without correcting for large scale dust emission. \textit{Right panel:} 
The same field after correcting for this emission as discussed in Section \ref{sec:dust}. The maps are smoothed with a beam of FWHM = 5$\arcmin$ for display purposes. }
\label{fig:dust}
\end{centering}
\end{figure}

The nature of the Galactic emission is not uniform across the full SPT-SZ survey. In particular, in areas of high dust intensity---such as are found at the western edges of the SPT-SZ fields closest to the Galaxy---this simple cubic polynomial correction breaks down.  
This is expected, as previous works have shown that regions with bright cirrus also have significant contribution from dust emission associated with molecular gas which follows a different modified black body function than regions correlated with HI \citep[see discussion in e.g.,][]{planck11-24}.  
Given the relative weights of the frequency channels in the map construction, such incomplete subtraction of dust emission results in a reduction in the recovered tSZ signal  at large-scales in very dusty regions.  
While this bias will not affect analyses conducted on smaller-scale features (e.g., cluster detection at $z>0.25$, tSZ profiles on scales smaller than the emission, etc.) we do include pixel masks encompassing regions of poorer dust subtraction for those who wish to exclude these regions from their analysis. These masks are constructed from the smoothed 545 GHz  foreground maps in regions with emission $>1.1$ MJy/sr  for which the residuals of the lower frequency dust subtracted foregrounds maps are $>3\sigma$ away from zero. As a check of our dust removal,  we cross-correlate our minimum-variance and CIB/CMB-nulled Compton-$y$ maps with the HI map provided with the HI4PI survey finding the cross-correlation power reduced by 7-10$\times$, respectively following this dust removal procedure. 
In Figure \ref{fig:dust} we show one SPT field, \texttt{ra5h30dec-55}, before and after removing Galactic dust. 

\subsection{Treatment of Missing Modes}\label{sec:trough}

As mentioned in Section~\ref{sec:mapmaking}, the SPT scanning strategy and data filtering results in a set of 
two-dimensional Fourier modes on the sky being removed from the SPT-SZ maps used here. In particular, the 
polynomial filtering of the SPT-SZ detector time-ordered data removes modes at low frequency along the scan
direction. For nearly all observations, SPT is scanned along lines of constant elevation, which, at the geographic
South Pole, correspond to lines of constant declination. In the Sanson-Flamsteed projection used in this work, 
lines of constant declination are parallel to the $x$-axis; thus, the SPT-SZ maps used here have no information
at the lowest values of $l_x$, or angular frequency along the horizontal map direction. \planck\ data is not missing 
these modes, so the LC algorithm uses \planck\ data exclusively in this region of Fourier space. At high 
$\ell = | \mathbf{l} |$, the \planck\ data is heavily rolled off by the finite angular resolution of the instrument, and 
the factor needed to restore these modes to an unbiased estimate of CMB/kSZ or Compton-$y$ multiplies the \planck\ noise, 
resulting in the high-$l_y$, low-$l_x$ part of Fourier space being many times noisier than average. 

We have chosen
to suppress this noise by applying a two-dimensional Fourier-space filter to the output component maps. This
filter is equal to unity everywhere except in the ``trough'' of modes at low $l_x$---in this trough the filter is equal 
to a Gaussian in $l_y$, with $\sigma_{l_y}$ set such that the noise is roughly equal to the noise outside the trough.
We include two-dimensional ($l_x$,$l_y$) and one-dimensional ($\ell$-space) versions of this filter
as one of the data products in our release; these can be used as an effective transfer function or $\ell$-space beam
in power spectrum or correlation analyses. For object-based analyses such as aperture photometry, we estimate
the bias from this filter in Section~\ref{sec:troughbias}.

\subsection{CIB Contamination vs.~Noise in ``Minimum-variance'' $y$ Map}\label{sec:herschelcross}

\begin{figure*}
\includegraphics[width=\columnwidth]{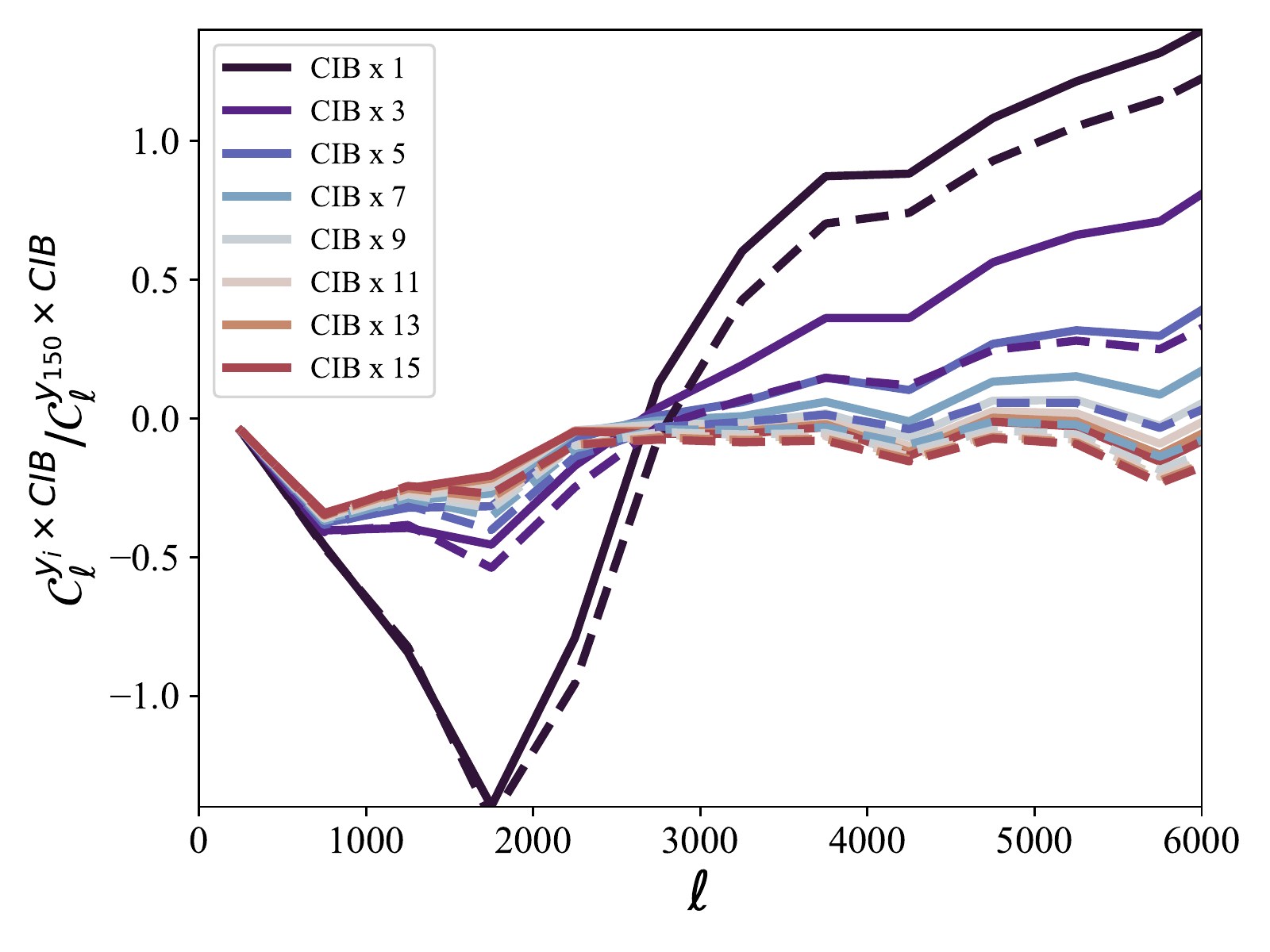} 
\includegraphics[width=\columnwidth]{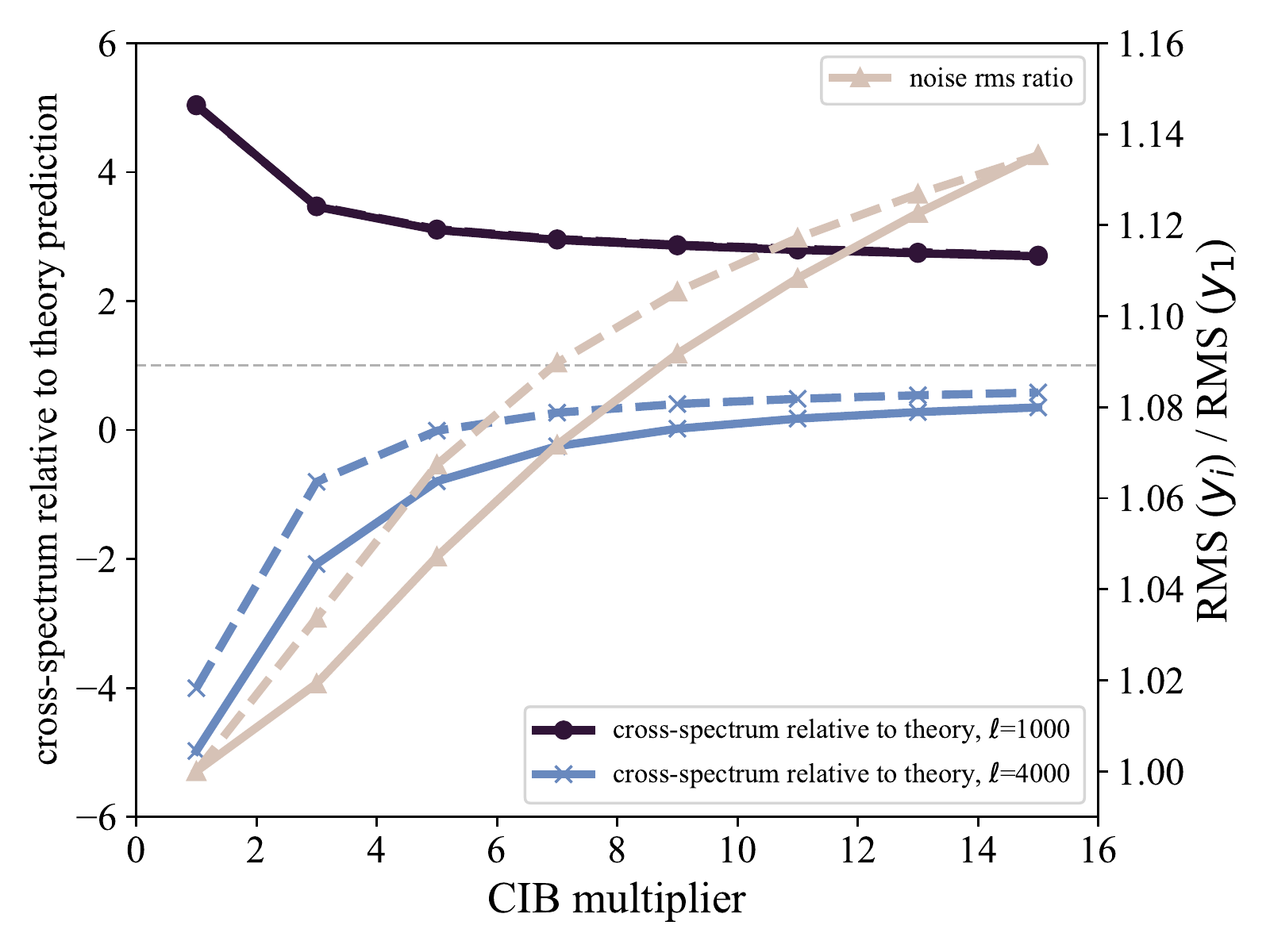} 
\caption{
\textit{Left panel:} 
$y$-CIB cross-spectrum for different CIB covariance multiplier values (see text for details) 
relative to the $y$-CIB cross-spectrum for a $y$ map made just from dividing a 150~GHz map
by the tSZ factor for that frequency. 
Solid lines show results using band weights created assuming
normal-depth SPT-SZ field noise; dashed lines show results using band weights created assuming
``double-depth'' SPT-SZ field noise.
\textit{Right panel:} 
$y$-CIB cross-spectrum ratio 
at $\ell=1000$ and $\ell=4000$ 
divided by the MDPL prediction for that ratio (see text for details) 
and total $y$-map rms as a function 
of CIB multiplier. 
As in the left panel, normal-depth and double-depth results are indicated with solid and 
dashed lines, respectively. (The single- and double-depth results for $\ell=1000$
are indistinguishable, because that multipole range is dominated by \planck\ data.)
From this plot it is clear that a significant reduction in spurious $y$-CIB correlation 
(caused by residual CIB in the synthesized $y$ map)
can be 
achieved with a small noise penalty.}
\label{fig:herschelcross}
\end{figure*}

As discussed above, the band weighting in the single-component version of the linear combiner (in which
no other signals are explicitly nulled) produces the minimum total variance in the resulting Compton-$y$ map.
But some sources of variance are of more concern than others, and for a $y$ map, the presence of CIB in 
the map is generally of the highest concern, particularly for using that $y$ map in cross-correlation studies. One 
can attempt to null a component of the CIB with a particular SED, as we do in the three-component version of the 
map we create, but that results in a fairly high noise penalty and does not fully solve the problem, as the
CIB is a combination of many components with different SEDs. Multiple components with different SEDs
can be nulled, but this is limited by the number of frequency bands (and further increases the noise penalty).

Here we investigate trading off CIB for 
noise and CMB variance by artificially increasing the weight of the CIB in the linear combiner---effectively
lying to the algorithm about how bright the CIB is. We find that multiplying the CIB part of the covariance
by some amount is quite effective in reducing the CIB component in the resulting $y$ map while only 
causing mild increases in total variance.\footnote{We note that this is effectively a less mathematically rigorous version of the Partially Constrained ILC method proposed in \citet{abylkairov20}.}
We measure the level of CIB in the synthesized ``minimum-variance'' $y$ map by 
cross-correlating the \textit{Herschel}/SPIRE 500$\mu$m map described in Section~\ref{sec:herschel} 
with the $y$ map from the SPT-SZ field that overlaps with the \textit{Herschel}/SPIRE coverage,  
keeping in mind that there should be some level of true, underlying correlation
between the $y$ and CIB fields \citep[e.g.,][]{george15,planck15-23}.

Figure~\ref{fig:herschelcross} shows the measured cross-correlation between the synthesized $y$ map 
and the \textit{Herschel}/SPIRE 500$\mu$m map as a function of CIB covariance multiplier. 
Specifically, the left panel of Figure~\ref{fig:herschelcross} shows the ratio of the measured 500$\um$-$y$ cross-spectrum
to the cross spectrum of the 500$\um$ map with a $y$ map made just by dividing an SPT-SZ-\planck\ synthesized 
150~GHz map by $\fsz(150\;\mathrm{GHz}) \tcmb$, with different curves representing different values of the CIB multiplier. 
The right panel shows the ratio of the measured 500$\um$-$y$ cross-spectrum to the predictions of the cross-spectrum between $y$ and
the CIB at 500$\um$ from the Omori et al.~(in preparation) $y$ and CIB skies painted onto the MDPL2 \citep{klypin16} simulations.
Curves are shown for two specific $\ell$ values ($\ell=1000$ and $\ell=4000$) as a function 
of CIB multiplier, along with the total map rms relative to the true minimum-variance $y$ map (in which
the ``CIB multiplier'' is unity).
In both plots, different curves are shown for band weights synthesized assuming
the noise in the normal SPT-SZ field depths and for band weights synthesized assuming the noise in the 
``double-deep'' fields. 

It is clear that the contribution to the 500$\um$-$y$ cross spectrum from residual CIB in the 
synthesized $y$ maps can be reduced significantly 
with only a mild ($\lesssim 10\%$) noise rms penalty. 
Using a by-eye metric of where the cross-spectrum appears to asymptote at both ell values, 
we choose a CIB multiplier value of eight
for the normal-depth fields and five for the double-depth 
fields.
When we use the term ``minimum-variance map'' anytime in the rest of this work, we are actually
referring to these maps. Finally we note that we do not apply this CIB multiplier to the covariance when we
produce $y$ or CMB maps with a component of the CIB formally nulled.

\section{The CMB/kSZ and Compton-$y$ Maps}\label{sec:mapproducts}

\begin{figure*}[ht]
\begin{center}
\vspace{0.3cm} 
\includegraphics[width=7.25in, trim=0cm 0 0 15 mm]{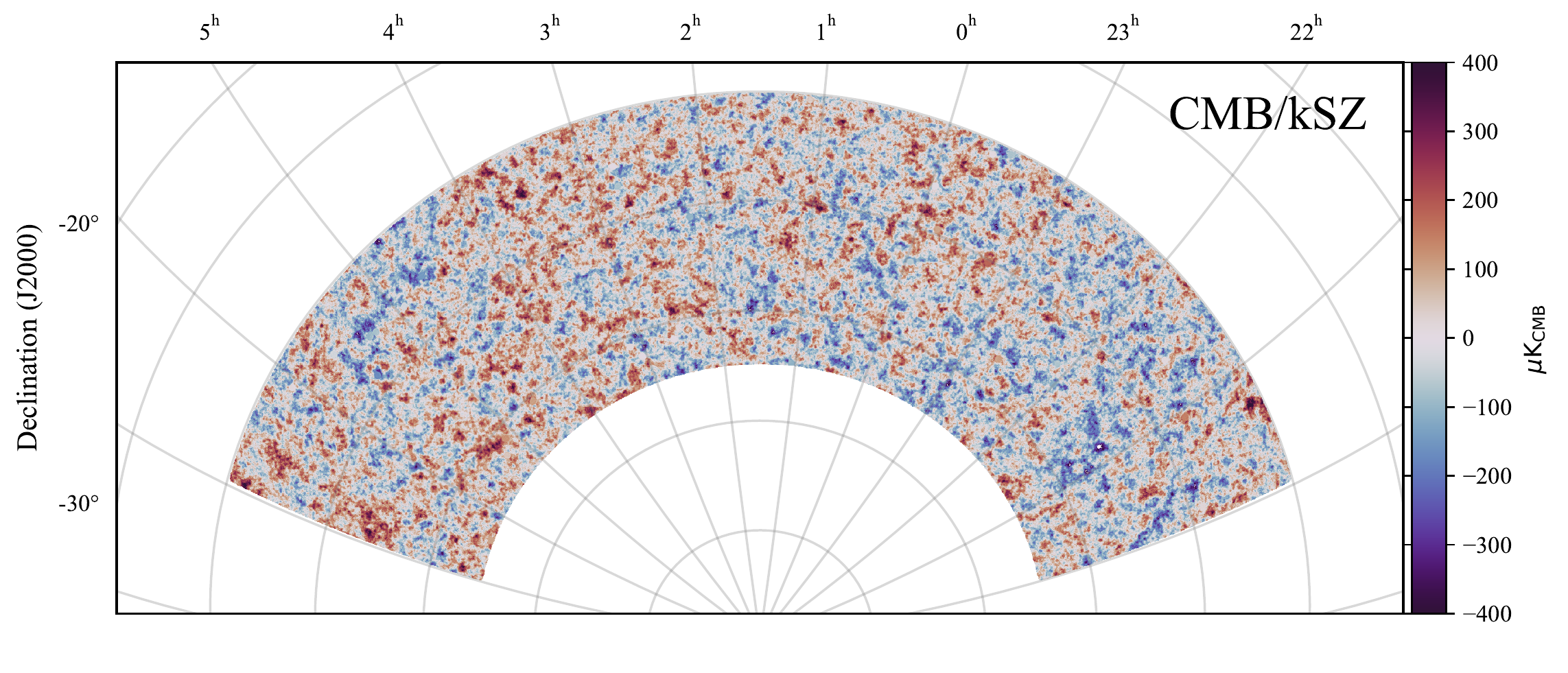} 
\includegraphics[width=7.16in, trim=0cm 0 0 20 mm]{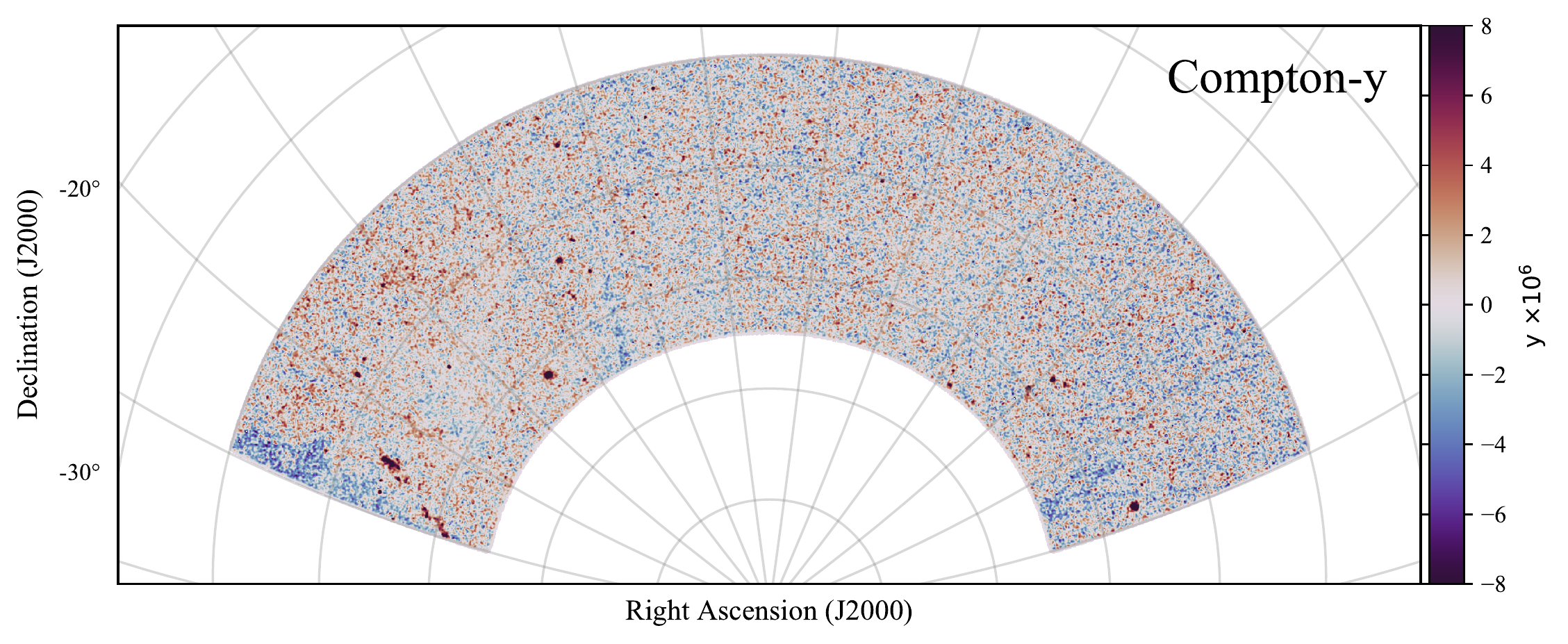} 
\end{center}
\caption{\label{fig:mapfigure} Two examples of component maps provided in association with this work. \textit{(Top)} tSZ-nulled CMB/kSZ map constructed using observations from the 2500d SPT-SZ survey and  \planck \ HFI. \textit{(Bottom)} Minimum-variance Compton-$y$ map from the same. Both of these maps are smoothed here for display purposes; we provide minimum-variance map products at  $1\farcm25$ resolution and map products with other signals nulled at  $1\farcm75$ resolution in the data release. }
\end{figure*}

We make and release several versions of component maps resulting from our LC algorithm:
\begin{itemize}
\item a ``minimum-variance'' $y$ map. This product is constructed using band weights defined 
in Equations~\ref{eqn:ymap}--\ref{eqn:sigmapsi}. Assuming the signal and noise models are correct, 
and with no modifications to the input covariance, this product formally has the lowest variance
possible for an unbiased estimate of $y$ on all scales. However, as discussed above, we do 
modify the covariance to reduce the correlation of this map with the CIB (as traced by 
500$\mu$m \textit{Herschel}/SPIRE data), at the cost of a small (order 10\%) penalty in total $y$-map rms.
We nevertheless will refer to this map as the ``minimum-variance' map" (or ``single-component map") 
in the remainder of this work.
\item a ``CMB-nulled" $y$ map and the corresponding $y$-nulled CMB/kSZ map. We also refer to these
as the ``two-component" maps.
\item a ``CMB- and CIB-nulled" $y$ map and the corresponding $y$- and-CIB-nulled CMB/kSZ map and 
$y$- and CMB-nulled CIB map. We also refer to these as the ``three-component" maps.
Because the CIB is made up of many independent emitters at different redshifts and temperatures
(and different compositions), it is not possible to null all the CIB emission with a single SED.
For our three-component maps, we choose to null CIB emission with an SED equal to the best-fit
Poisson-component SED in \citet{george15}, who find that this component is the largest contribution
to the $\ell=3000$ power at 150~GHz.
As expected, the noise levels of the two- and three-component products are significantly higher than for the 
minimum-variance case.
\end{itemize}
As an example of two of the data products we provide, we show the full-survey versions of the  tSZ-nulled CMB/kSZ  and the minimum-variance Compton-$y$ maps in Figure \ref{fig:mapfigure}. 
We provide flat-sky full and half (survey+mission) maps in each of the 19 SPT fields at $0\farcm25$ arcminute pixelation as well as combined maps covering the full SPT-SZ footprint in the HEALPix format with $N_{\mathrm{side}}=8192$. 
The minimum-variance single-component data products are provided at $1\farcm25$ resolution and the multi-component products at  $1\farcm75$ resolution.

Figure~\ref{fig:weights} shows the relative band weights ($\psi$ from Equation~\ref{eqn:psi} or 
$\psi_i$ from Equation~\ref{eqn:psimult}) as a function of $\ell = |\mathbf{l}|$ for three 
of the maps discussed above. As expected, the \planck\ bands provide the bulk of the information 
at low $\ell$ (large angular scales), where the SPT-SZ data is contaminated by atmospheric fluctuations.
At higher $\ell$ (smaller angular scales), the higher angular resolution and lower instrument noise
levels of the SPT-SZ data take over. Also notable is the similarity of the single-component and two-component
tSZ weights at low $\ell$, understandable as the primary CMB is the largest source of variance at 
these scales. The tSZ-nulled CMB/kSZ weights are dominated at low $\ell$ by the \planck\ 217~GHz
data, which is close to the tSZ null, but at high $\ell$ all three SPT-SZ bands contribute, in part because 
of the higher relative noise in the SPT-SZ 220~GHz data.

\subsection{Bias to Aperture Photometry From Missing Modes}\label{sec:troughbias}

As discussed in Section~\ref{sec:trough}, the maps produced in this work have two filters
applied after the beam-and-transfer-function-deconvolved individual frequency maps have
been combined into synthesized component maps. One filter
is a simple Gaussian smoothing with full width at half maximum of $1\farcm{25}$ for the 
single-component $y$ map and $1\farcm{75}$  for the multi-component maps. 
The non-standard filter is an anisotropic low-pass filter that only affects modes
that oscillate slowly in the direction of right ascension. The motivation for this filter is discussed
in Section~\ref{sec:trough}; here we calculate the bias incurred to aperture photometry if this
filter is ignored. 

We estimate this bias by creating simulated galaxy cluster profiles, Fourier transforming, 
multiplying by this $\mathbf{l}$-space filter, inverse-Fourier transforming, and performing aperture
photometry on these simulated profiles. We compare the results of aperture photometry on 
the filtered profiles to results of aperture photometry on the original cluster profiles and report
the difference as the bias from the filter. We perform this calculation on a range of cluster masses
and redshifts and using a range of aperture radii. We use the \citet{arnaud10} model for the 
cluster $y$ profile. 
For cluster masses $0.5 \times 10^{14} M_\odot < M < 10 \times 10^{14} M_\odot$, redshifts
$0.1 < z < 2$, and aperture radii $1.0^\prime < r < 10.0^\prime$, we find a maximum bias of 
$\sim$3\%. The bias is larger at smaller apertures (at $r > 5^\prime$ the maximum bias is 
$\sim$1\%).

\subsection{Comparison to Other Releases}\label{sec:compare}

\begin{figure}[t]
\begin{centering}
\includegraphics[width=\columnwidth]{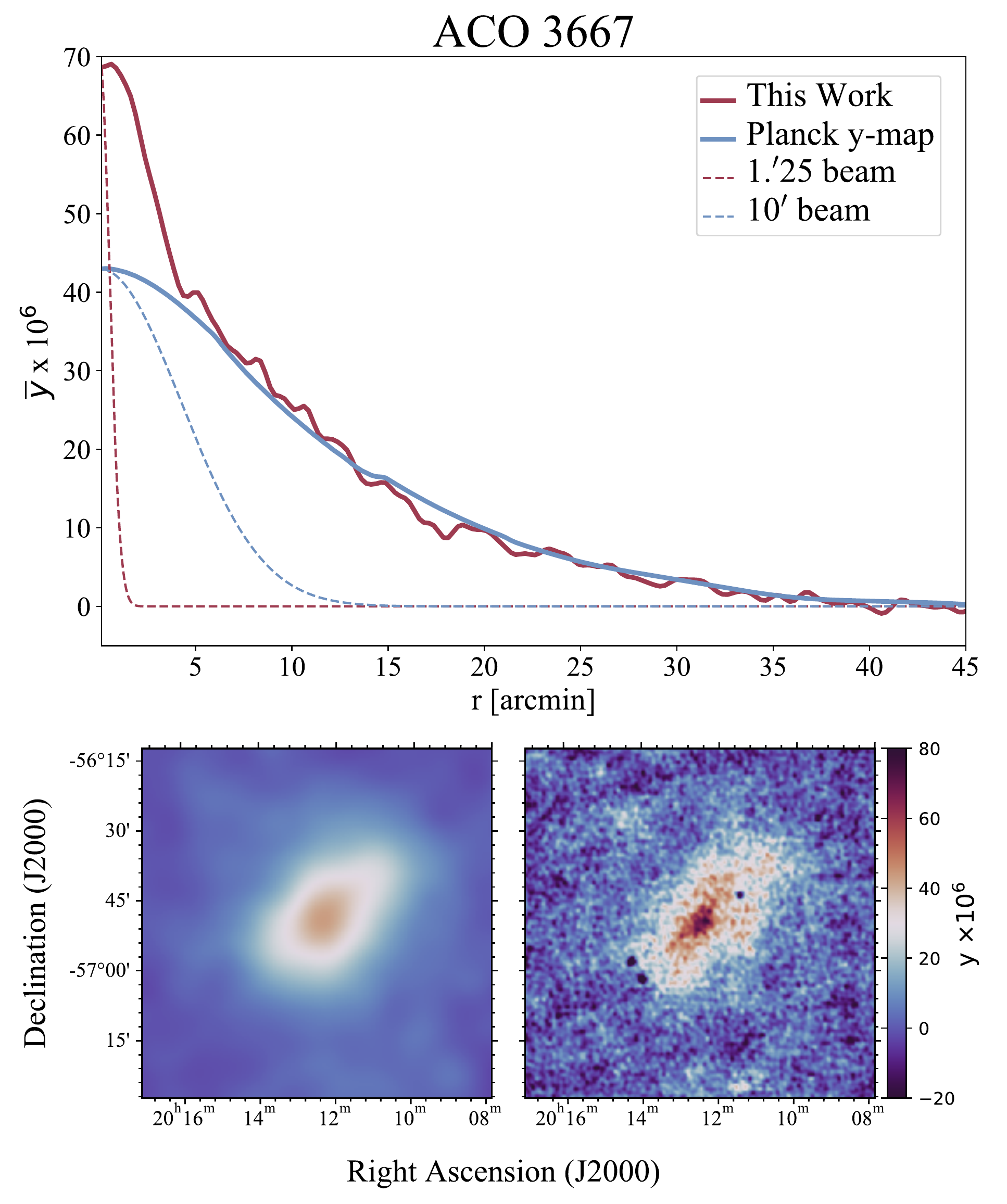} 
\caption{
\textit{Top panel:} Azimuthally averaged Compton-$y$ profiles for ACO 3667 at $z=0.0556$.  
\textit{Bottom panels}: On the left is a cutout from the \planck \ Compton-$y$ map \citep{planck15-22} centered on the cluster. We make a similar cutout on the right from the minimum-variance Compton-$y$ map presented in this work. The arcminute-scale circular decrements in this image are located at the locations of bright radio sources readily detected in SPT data, enabling us to mask these regions when measuring the cluster's pressure profile.}
\label{fig:cluster_profile}
\end{centering}
\end{figure}

To date, the only other publicly available Compton-$y$ maps in the SPT-SZ survey footprint are derived from \planck \ data products (particularly \citealt{planck15-22}).
As expected, the addition of SPT data products significantly improves both the angular resolution (from $10\arcmin$ to $1\farcm25$ or $1\farcm75$, for the minimum variance and component-nulled versions, respectively) as well as the noise performance at small scales. 
We provide a qualitative demonstration of this effect in Figure \ref{fig:cluster_profile} where we highlight the ability of the new data products to constrain arcminute-scale features in the pressure profile of Abell~3667 \citep{duus77, abell89} at redshift $z=0.0556$ \citep{struble99} for which the improved angular resolution allows us to resolve features on  $\sim80$ kpc scales.
This significant improvement in resolution will also aid in the detection of higher-redshift clusters; we detail the results of a blind cluster search below in Section \ref{subsec:clustercheck}. 

\begin{figure*}
\begin{centering}
\includegraphics[width=\columnwidth]{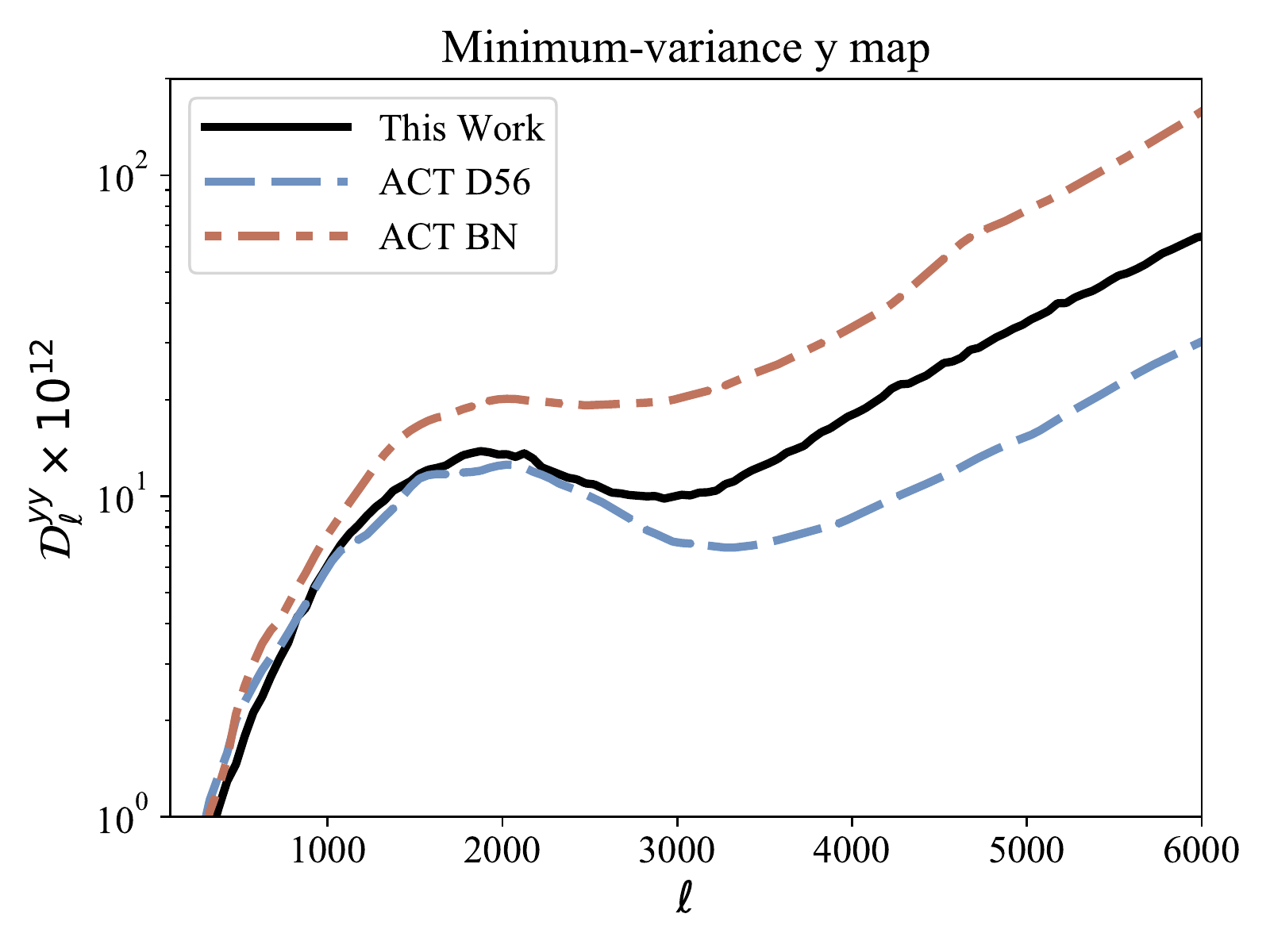} 
\includegraphics[width=\columnwidth]{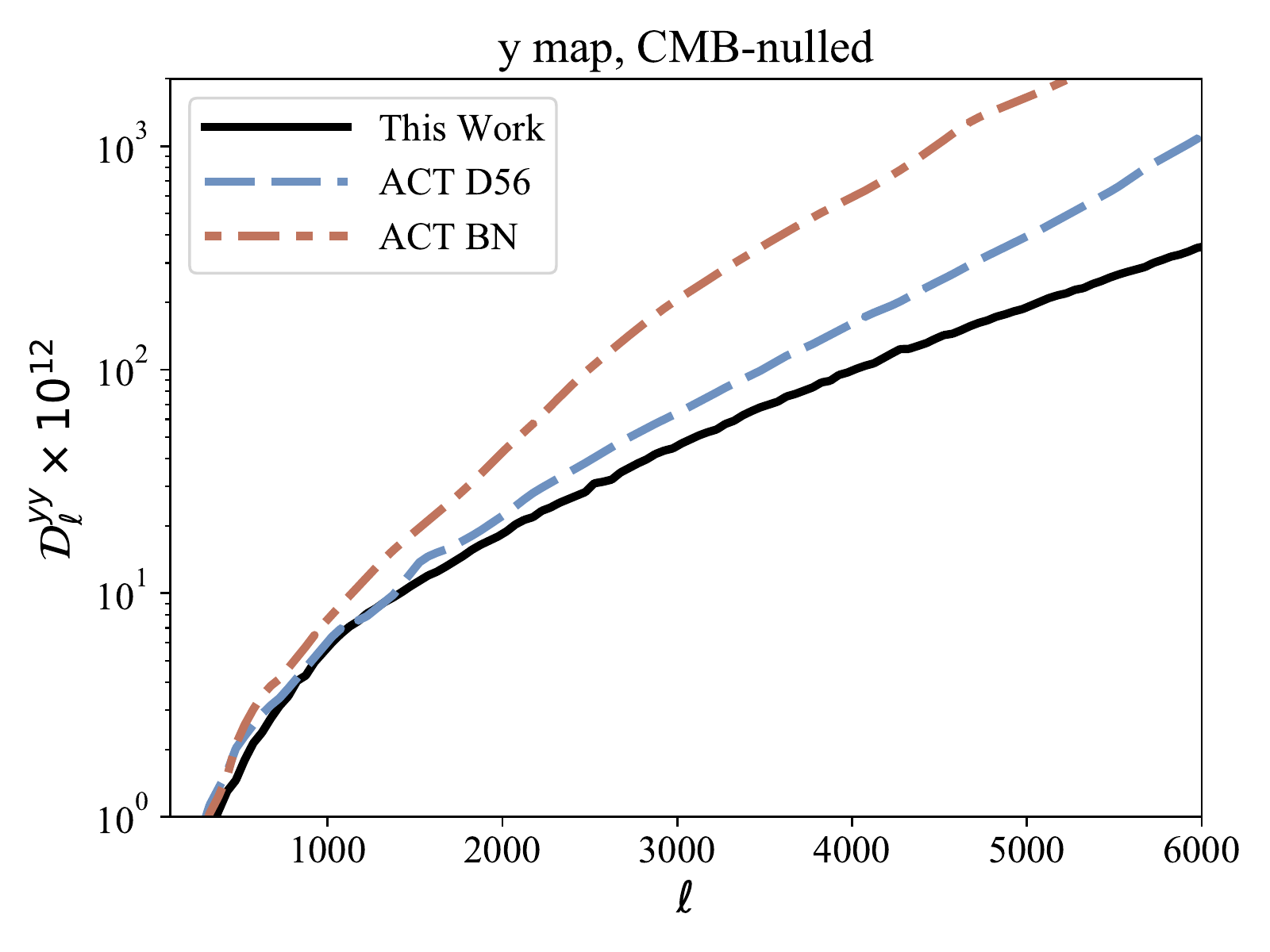} 
\includegraphics[width=\columnwidth]{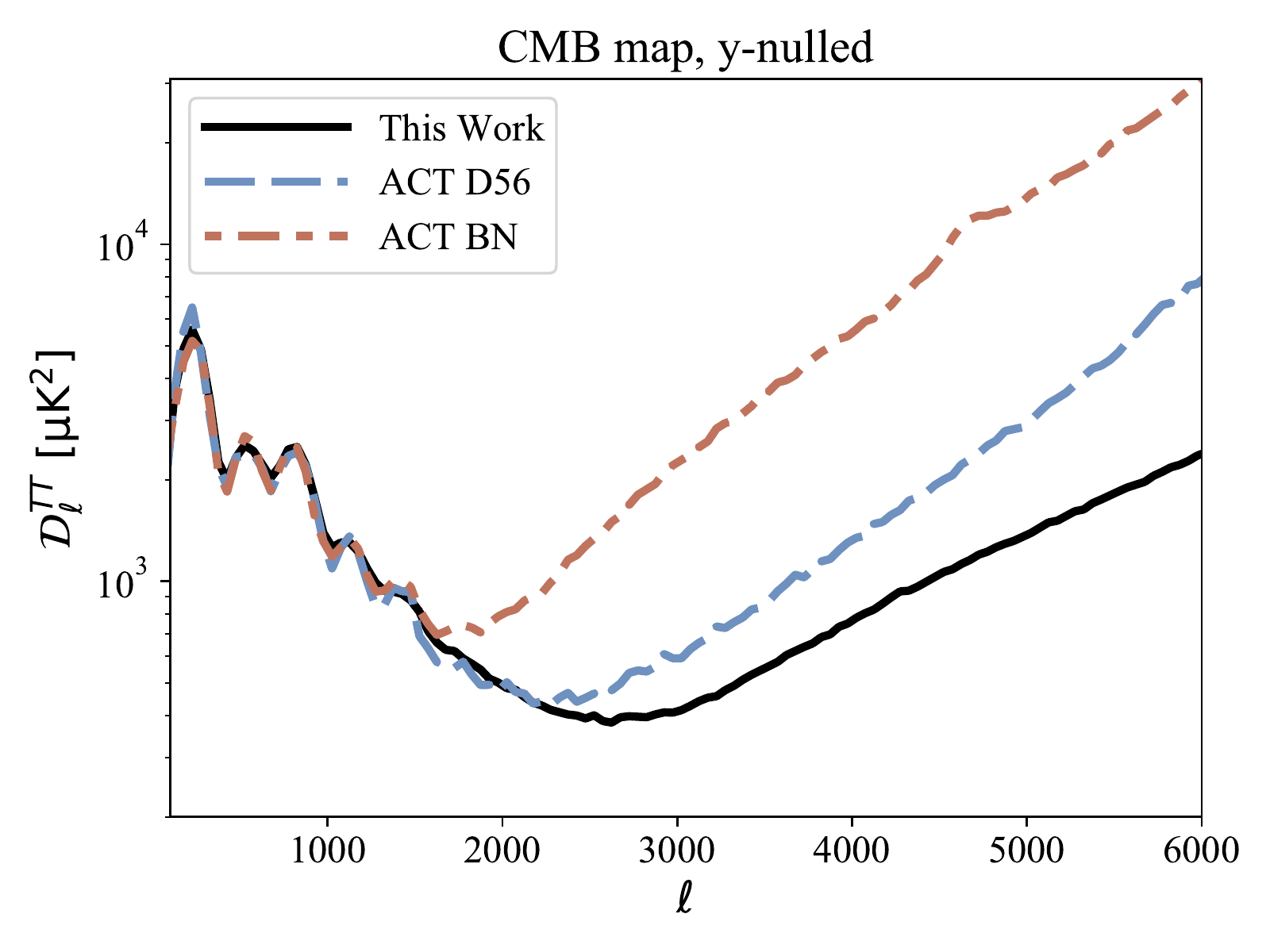} 
\includegraphics[width=\columnwidth]{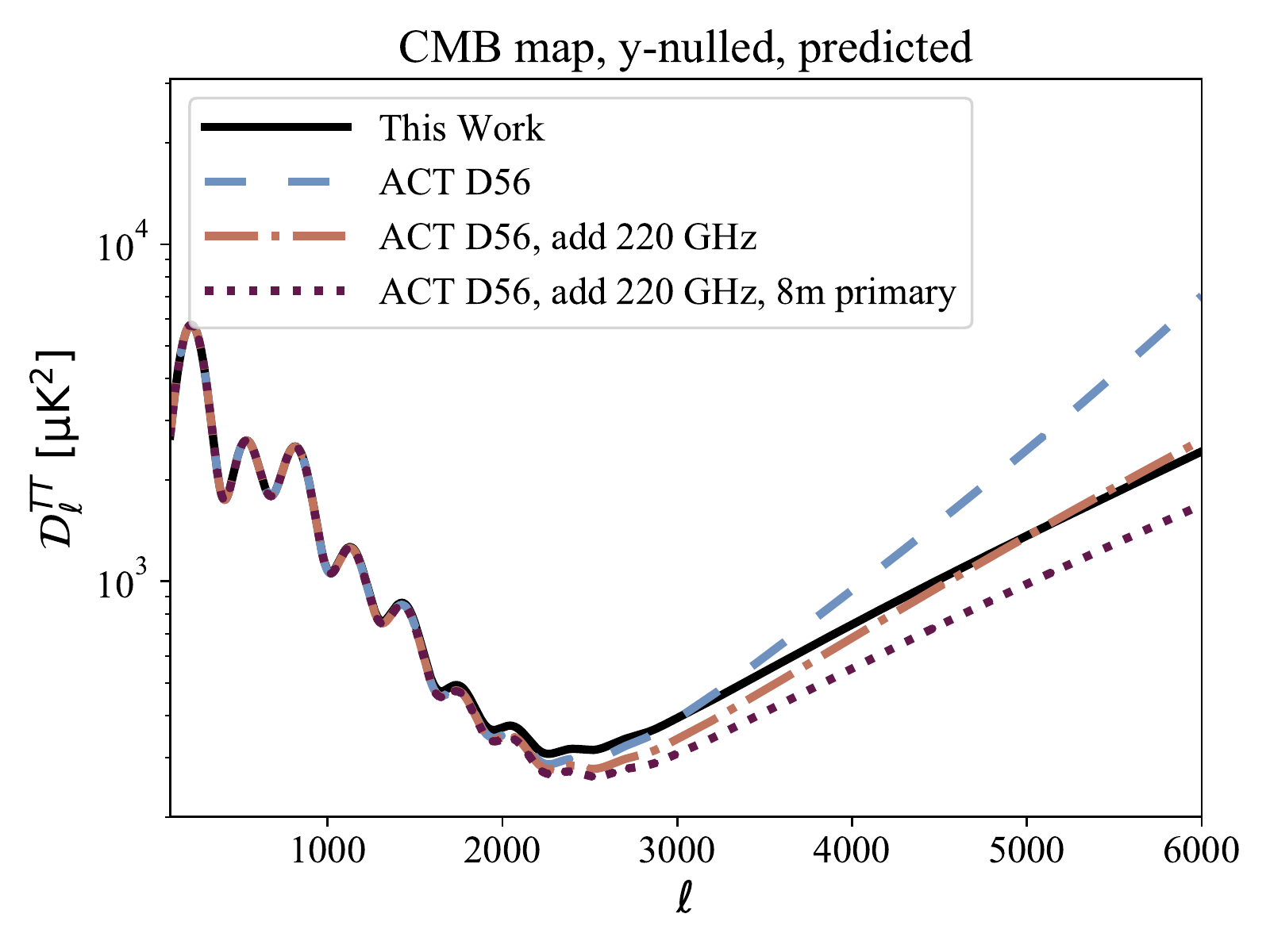} 
\caption{Angular power spectra of selected output maps, with results from ACT+\planck\ \citep{madhavacheril20} shown for comparison.
The agreement between the three curves in the signal-dominated multipole region 
($\ell \lesssim 2200$ for SPT and ACT D56, $\ell \lesssim 1500$ for ACT BN) 
provides a rough
validation of both sets of data products. The differences in the noise-dominated multipole region
are discussed in the text.
\textit{Top left}: Power spectra of the minimum-variance $y$ maps from this work and from the two 
different sky regions used in \citet{madhavacheril20}.
\textit{Top right}: Similar for the CMB-nulled $y$ map.
\textit{Bottom left}: Similar for the $y$-nulled CMB/kSZ map.
\textit{Bottom right}: Predicted power spectra for the $y$-nulled CMB/kSZ map from  
SPT and the ACT D56 region, using noise levels reported in
this work and Table 1 of \citet{madhavacheril20}. To demonstrate the importance of high-frequency data
and angular resolution in the component-separation process, 
curves for ACT D56 noise levels are shown in the hypothetical
scenario in which 220~GHz data (at the rough noise levels of the SPT data) are added and in which the resolution
of the ACT data is improved (see text for details).}
\label{fig:autospectra}
\end{centering}
\end{figure*}

As mentioned in the introduction, arcminute-scale component maps---including minimum-variance and CMB-nulled 
Compton-$y$ maps and a $y$-nulled CMB/kSZ map---constructed from \planck\ and ACT data were presented in 
\citet{madhavacheril20}. In Figure 9 of that work, the authors showed that the power spectra of the $y$ and CMB/kSZ maps
including ACT data were (as expected) signal-dominated to much higher multipole values (smaller scales) than 
corresponding maps from \planck\ alone. In turn, we compare the power spectra of the maps in this release to those
from the two sky regions in \citet{madhavacheril20}---the $\sim$1600-deg$^2$ BOSS North or BN field and a deeper 
$\sim$500-deg$^2$ field referred to as D56---in Figure~\ref{fig:autospectra}.

There are many features worth noting in Figure~\ref{fig:autospectra}. First, the agreement in the signal-dominated
multipole regions---in particular of the CMB/kSZ map (bottom left panel)---demonstrates that the output maps from the
two works are statistically consistent at the 10\% level or better. This is not surprising, especially because the maps from
both works are dominated by \planck\ data at these angular scales, but it is a rough validation of both pipelines. 

The amplitudes of the power spectra in the noise-dominated multipoles merit further discussion. As shown in 
Figure~\ref{fig:weights} in this work 
and Figure 5 in \citet{madhavacheril20}, the high-$\ell$ ($\ell \gtrsim 3000$) ``minimum-variance'' $y$ map weights are dominated
by the 150~GHz SPT / 148~GHz ACT data. The typical SPT-SZ noise level at 150~GHz ($\sim$16 $\mu$K-arcmin) is 
between the 148~GHz noise levels in the two ACT regions ($\sim$25 and $\sim$10 $\mu$K-arcmin in BN and D56, respectively, 
cf.~Table 1 in \citealt{madhavacheril20}), so it is not surprising that the high-$\ell$ power spectrum of the SPT-SZ 
``minimum-variance'' $y$ map lies in between that of the two ACT regions. By contrast, the SPT-SZ high-$\ell$ 
power spectrum is below that of both ACT regions for both components of the two-component linear combination
(the CMB-nulled $y$ map and the $y$-nulled CMB/kSZ map). This is especially surprising given that the 
98~GHz ACT noise levels in the D56 region ($\sim$17$ \mu$K-arcmin) are significantly lower than the SPT-SZ 95~GHz noise. 

The two features
of the data used in the SPT-SZ linear combination that could account for this are: 1) data from the SPT-SZ 220~GHz
band are included; 2) the SPT-SZ data are at slightly higher resolution, owing to the 10-meter SPT primary mirror
compared to the 6-meter ACT primary (though the diameter of the primary region illuminated by the SPT-SZ 
camera is closer to eight meters). In the bottom-right panel of Figure~\ref{fig:autospectra}, we show that both of 
these features contribute to the lower SPT-SZ noise in the two-component maps. The dot-dashed line shows 
the predicted ACT-D56 $y$-nulled CMB/kSZ power spectrum if 220~GHz data (with roughly the SPT-SZ noise 
level) are added, and the dotted line shows that power spectrum if 220~GHz data are added and if the resolution
in all bands is improved by a factor of 1.33 (the ratio of the diffraction limits of 6-meter and 8-meter apertures).
This result demonstrates the utility of higher-frequency data and higher angular resolution in component separation,
particularly for the goal of extracting the tSZ-nulled kSZ signal out to $\ell \sim 4000.$

\section{Validation of the Component Maps}\label{sec:char}

In this section we detail several characterization analyses of the SPT-SZ + \planck \ component maps
described in the previous section.
As discussed in Section~\ref{sec:compare}, a rough validation of the maps, particularly the $y$-nulled CMB/kSZ
map is provided by comparing the power spectrum of maps from this analysis to the power spectrum of ACT+\planck\ maps in 
Figure~\ref{fig:autospectra}; we apply more quantitative measures in this section.
These tests include validation of our assumptions for the input signal model, tests of the blind detection from SZ clusters in the maps, and finally, a cross check of our tSZ nulling in the CMB/kSZ map products.

\subsection{Validation of the Input Signal Model}\label{sec:residual_contamination}

\begin{figure}
\includegraphics[width=1.0 \columnwidth, trim=15mm 8mm 0 0]{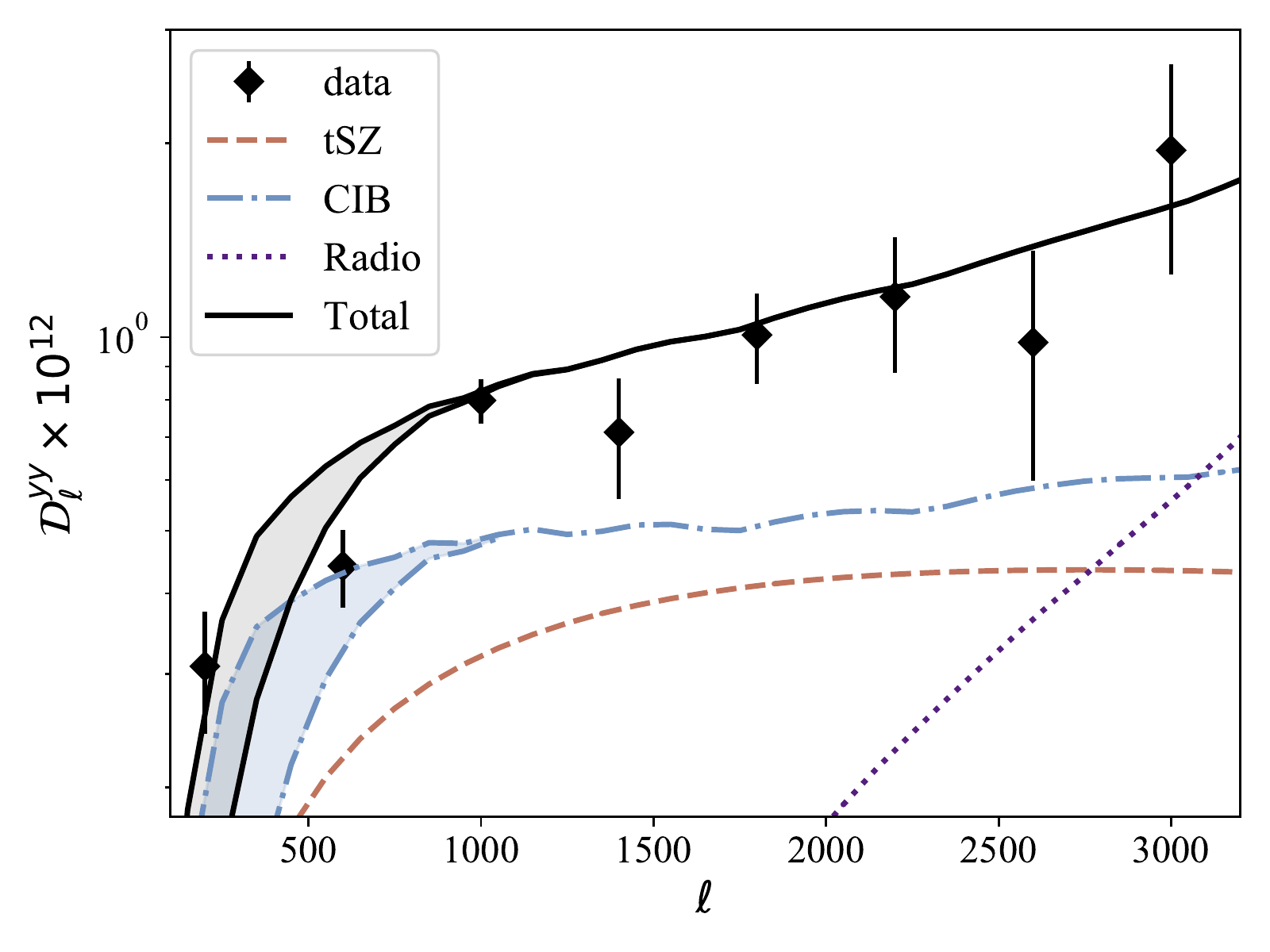} 
\caption{Power spectrum of the CMB-CIB-nulled $y$ map. The data points
show the measured power spectrum of the $y$ map from the three-component
analysis in this work. The power spectrum is estimated using independent halves
of the data to avoid noise bias. 
Error bars are estimated using resampling of the values from the individual SPT-SZ fields.
Lines show the expectation for power from various components in the $y$ map, 
given the signal models used in the 
linear combination algorithm. These are calculated by multiplying the model power 
spectra at each band by the band weights for the CMB-CIB-nulled $y$ map and summing. 
The blue and gray shaded areas indicate the possible effect of dust cleaning on the CIB 
(see text for details).
The by-eye agreement between the solid black line (sum of all model components)
and the data points indicates that the signal variance models are roughly representative
of the true signals in the data.}
\label{fig:powerspectrum}
\end{figure}

As discussed in Section~\ref{sec:ymap}, the linear combination algorithm we use relies on models 
of the power in various sky components, including the cross-power between different frequencies.
Even if these models are wrong, the resulting CMB/kSZ or Compton-$y$ map (or other component maps) will still 
have unbiased response to the desired signal (provided the instrument bandpasses are correctly measured),
but the linear combination of bands used may not be optimal; i.e., there can be excess variance induced
by the incorrect modeling of sky signal. 

Figure~\ref{fig:powerspectrum} shows a simple check of the
signal modeling, namely the power spectrum of the output $y$ map in the case where the CMB/kSZ
and one component of the CIB are explicitly nulled. 
The data points are an average over power spectra from the 19 individual SPT-SZ fields, calculated using the
cross-spectrum of one half of the data against the other to eliminate noise bias. The data points are corrected 
by the $1\farcm{75}$ smoothing beam and the ``trough filter'' (see Section~\ref{sec:trough}) applied to the maps, but no other 
corrections are made (e.g., for mode coupling). Error bars are calculated using resampling (with replacement)
of the values from the individual fields.
Overplotted are the expected power from various
sky signals given the input model and the real band weights $\psi_i(\nu_j)$ returned by the linear combiner.
The model for the tSZ is the best-fit model from \citet{reichardt20}.
The subtraction of the dust template described in Section~\ref{sec:dust}, constructed by smoothing the 
\planck\ COMMANDER 545~GHz foreground map to $16^\prime$ resolution, will affect the low-$\ell$ CIB power in the output $y$ map; 
for the purposes of this plot we bracket the possible effect by multiplying the residual CIB power in the model
by a  $16^\prime$-FWHM Gaussian, subtracting that from the model, and plotting the result as a second set
of model lines. The shaded area between the unmodified and modified models thus indicates the modeling 
uncertainty introduced by the dust-cleaning step.
Because of this uncertainty and the lack of correction for, e.g., mode coupling, we do not report a formal
goodness of fit to the model; we simply note that at the 10\% level the total power in the model appears to 
agree with the data, particularly above $\ell = 1000$.

Somewhat more formally, we use the linear combination framework to investigate the potential impact on 
$y$-map variance from
incorrect signal modeling. The signals in question have been reasonably well characterized in previous 
studies. The primary CMB temperature anisotropy has been measured at the percent level in bins of 
$\Delta \ell \sim 50$ (assuming a smooth power spectrum) by \planck\ \citep{planck15-11} 
and high-resolution ground-based 
experiments such as ACT \citep{choi20} and SPT \citep{story13}. 
At $\ell = 3000$, \citet{reichardt20} measures the relevant secondary and foreground signals, 
namely the CIB and residual radio galaxy power after masking, with $\sim$10\% precision. We create 
several sets of band weights using different signal model inputs, modifying each input component by 
an amount that is in mild tension with current data. For each of these sets of band weights, we compute
the predicted variance in the output maps under the assumption that the modified inputs are correct 
and under the assumption that the original signal model was correct. 

Specifically, we modify the primary CMB power spectrum by $\pm 2\%$ in amplitude and $\pm 2\%$
in spectral index, both of which are four to five times the marginalized error bar on the associated 
cosmological parameter or parameter combination ($A_s e^{-2 \tau}$ and $n_s$, respectively).
We modify the amplitude of the input CIB model by $\pm 20\%$, and we double the model radio
power. We run the test 
described above in the one-, two-, and three-component cases. In all of these 
permutations, we find that the fractional excess variance from assuming the wrong model is never 
greater than $0.6\%$. 
We conclude that, for the noise levels and the size of the sky patch in this analysis, 
the sub-optimality of using models for the sky signal components is minimal.

\subsection{Detection of Clusters by the tSZ Effect}\label{subsec:clustercheck}
One natural test of the quality of a constructed Compton-$y$ map is of its efficacy for the blind detection of clusters via the tSZ effect.
As noted in the Introduction, there have been several recent analyses undertaking cluster searches on combined \planck \ and ground-based data using ACT \citep{aghanim19} and SPT-SZ data \citep{melin20}.  
These works have both noted significant improvement in cluster detection over data from each sample alone with \citet{melin20} particularly quantifying such improvements as a function of redshift. 
Here we will present the results of such a cluster search on our minimum-variance Compton-$y$ map, comparing our results  to cluster samples reported by the SPT-SZ \citep{bleem15b} and Planck collaborations \citep{planck15-27}.

To conduct this search, we adopt an essentially identical procedure to that used in previous searches for clusters in SPT data  (see e.g., \citealt{bleem15b} for a more detailed description of the process). Our search is based on the spatial-spectral filtering method presented in \citet{melin06} and,    
as the $y$-map construction has already isolated the tSZ signal,  we simply make use of a spatial filter designed to optimally extract the cluster signal in the presence of noise.  
We adopt a projected spherical $\beta$-model with $\beta$ fixed to 1 \citep{cavaliere76} as the model of our cluster profile: 
\begin{equation}
\Delta T = \Delta T_{0}(1+ \theta^{2}/\theta_\mathrm{c}^{2})^{-(3\beta-1)/2}
\label{eqn:beta}
\end{equation}
where the normalization $\Delta T_{0}$ is a free parameter and the core radius, $\theta_\mathrm{c}$, is allowed to vary in twelve equally spaced steps from  $0\farcm25$ to 3\arcmin \ (as in previous SPT works) and additionally include $\theta_\mathrm{c}$ values of $5\arcmin,7\arcmin,9\arcmin$, and $12\arcmin$ to allow for the detection of clusters of larger angular extent (the latter now enabled by the inclusion of the \planck \ data). 
 This filtering corresponds to scales of $\theta_\mathrm{500c}$ of approximately $1.25\arcmin$ to $60\arcmin$ in the commonly used \citet{arnaud10} model of cluster pressure profiles. 

We run our cluster identification algorithm on the flat-sky version of each of the 19 SPT-SZ fields, adopting a $1\farcm25$ beam as our transfer function.  
The noise in these maps is composed of the astrophysical, instrumental, and atmospheric noise that remains after the map combination process.  
We estimate this noise by summing the power from each component of our sky + noise model (Section \ref{sec:skymodel}) by the appropriate 2D Fourier space weights for each frequency map included in the $y$ map. 
We also heavily penalize the noise at low $k_x$, high $k_y$ where we are missing modes in the maps (Section \ref{sec:trough}). 

The resulting SZ candidate list has 418 detections at signal-to-noise $\xi>5$ (compared to 423 in \citealt{bleem15b}, but detected in 3\% less area, see Section \ref{sec:mapmaking})  and 677 $\xi>4.5$ (vs. 696). Fifty-five new SZ cluster candidates at $\xi>5$ (the 95\% expected purity threshold) are identified that are not in the \citet{bleem15b} sample. 
Recognizing there will be scatter between detection significances for given clusters between the lists owing to the inclusion of the \planck \ data, the addition of radio power in the noise covariance matrix used in the map construction (this power was not included in the \citealt{bleem15b} noise model),  as well as the additional masking around bright sources (Section \ref{sec:inpaint}), we match our new candidate list at $\xi>4$ to the published SPT-SZ list at $\xi>4.5$. 
Associating detections within $2\arcmin$ we recover 473 matches with an average ratio in the detection significance ($\xi_\textrm{new}/\xi_\textrm{SPT-SZ}$)=1.0 and standard deviation of this ratio of 0.2.  
We recover all confirmed SZ clusters \citep[those with redshifts reported in ][]{bleem15b} above $\xi=6.1$ within the unmasked Compton-$y$ footprint and 94\% (85\%) of confirmed clusters at $\xi>5$ ($\xi>4.5)$.  
Examining the locations of \cite{bleem15b} clusters not detected in our new search at $\xi>4$ in the signal-to-noise maps produced by the cluster finder we find a median difference of $\Delta\xi=1.4$; there are no significant redshift or $\theta_\mathrm{c}$ trends in the unmatched clusters compared to the matched. 

We note that a quick reading of this work may imply that the addition of \planck \ data did not contribute in a meaningful way for the detection of these systems. However, we remind the reader that we have adopted a weighting scheme in the map construction that minimizes the cross-correlation of the $y$ map with the CIB (see Section \ref{sec:herschelcross}) and adds a noise penalty to these maps compared to a true minimum-variance $y$ map. The addition of the \planck \ data (and to a much lesser extent the SPT-SZ 220~GHz data) essentially compensates for this choice.

Moving on to comparisons with the \planck \ SZ cluster sample \citep{planck15-26}, we associate our new SZ candidate list with confirmed \planck \ clusters within  the larger of $4\arcmin$ or the \planck \ positional uncertainty, finding 103/117 matches at $\xi>4$ and \planck \ detection signal-to-noise ratio $>4.5$ with a median improvement in the signal-to-noise of the detection of $1.25\times$. 
This is an improvement over a similar matching exercise undertaken with the \citet{bleem15b} sample for which 83 SPT-SZ clusters in the Compton-$y$ map footprint had matches to confirmed clusters in the \planck \ catalog (see also discussion comparing properties of the \planck \ and SPT-SZ and SPT-ECS samples in \citealt{bleem20}). The most significant gains were found, as would be expected,  for low-$z$ clusters with larger angular extent on the sky for which the inclusion of \planck \ data provides enhanced sensitivity. 
While the typical signal-to-noise  of the combined map SZ detections are higher than those from \planck \ alone, our simplified treatment of the \planck \ noise---both in the map combination procedure as well as the cluster extraction (see e.g., discussion in \citealt{planck13-20, planck15-26} for more optimal treatment of \planck \ data)---leads to lower detection significances for the lowest-redshift clusters. 

\subsection{tSZ nulling in the CMB/kSZ maps}

We perform one final validation test  regarding the efficiency of the tSZ nulling in the CMB/kSZ component maps. 
While the derivations governing both nulling and the thermal SZ spectrum are unambiguous, this test will probe both our constraints on the bandpasses and calibration of our maps as well as the presence of correlated signals not explicitly cancelled (from e.g.,  cluster member galaxies) at the locations of clusters; we note that as these are SZ-selected clusters this is not a strong test of the presence of such associated emission in the general cluster population \citep[though see discussion of such effects in e.g., Section 3.5 of ][]{bleem20}. 
As our baseline we construct a true minimum-variance CMB/kSZ map and sum the temperature values in this map at the location of 504 confirmed SPT-SZ clusters with $\xi>4.5$ that are not affected by  the masking used in this work. 
We find the mean temperature and its uncertainty at these cluster locations to be $-34\pm1.5 \mu$K. 
Performing this exercise on the CMB/kSZ maps with the tSZ- and tSZ/CIB-model nulled  we find  $-3\pm3 \mu$K and $7\pm7 \mu$K, respectively. 

Sharpening our sensitivity to cluster scales, we then repeat this exercise, performing compensated aperture photometry on the 3 maps with 2\arcmin \ apertures at the cluster locations. 
We again find a significant average temperature decrement in the minimum variance CMB map ($-1.16\pm0.025 \times 10^{-5}\mu$K-sr, 45$\sigma$), a small decrement in the tSZ-nulled maps ( $-3.1\pm 0.4 \times 10^{-6}\mu$K-sr; 7.5$\sigma$) and no significant detection in the tSZ/CIB-nulled maps ($3\pm2 \times 10^{-6}\mu$K-sr; 1.5$\sigma$ increment). 
As expected, the minimum-variance CMB map shows a significant temperature decrement at the stacked location of these massive systems in both tests and the  results from tSZ/CIB-nulled maps are consistent with zero. 
At first glance, the significant decrement in the stack in the tSZ-nulled map may indicate a problem in this map. However, we note that owing to noise bias in the tSZ selection we would not have expected complete cancellation in the tSZ-nulled CMB map. Recall the CIB is one of the dominant sources of astrophysical noise in the SPT maps at small scales \citep{reichardt20} and indeed a stack of the  SPT 220~GHz maps (whose effective frequency is at the tSZ null, Section \ref{sec:bandpasses})  shows a small, but significant, average temperature decrement demonstrating that the SPT-SZ cluster detection is slightly biased to areas with lower levels of CIB. Based on the weights in the CMB/kSZ two component map (Figure \ref{fig:weights}) we would expect a slight temperature decrement to remain at the location of SPT-selected clusters.

The best way to robustly test these maps is with a large cluster sample selected independently of any signal in the millimeter-wave maps. 
In lieu of such a sample, we perform a much more restricted test using optically selected clusters from the Blanco Cosmology Survey as presented in \citet{bleem15a}. 
Stacking a sample of 169 lower-mass clusters at $0.25 < z < 0.7$ with optical richness $\lambda >25$ we find a 7$\sigma$ temperature decrement  in the minimum-variance CMB maps ($-1.3\pm0.2 \times 10^{-6}\mu$K-sr) and no significant detection in the other two maps ($0\pm6 \times 10^{-7}\mu$K-sr and  $2\pm3 \times 10^{-6}\mu$K-sr in the tSZ- and tSZ/CIB-nulled maps respectively). Up to its limited statistical constraining power this test demonstrates the efficacy of tSZ removal in the tSZ-nulled map; we note that much more constraining tests will soon become possible with the release of eROSITA \citep{merloni12} and  Dark Energy Survey \citep{flaugher15} Year 3 cluster samples.

\section{Summary}\label{sec:summary}

In this work we have presented new CMB/kSZ and Compton-$y$ component maps from combined SPT-SZ and \planck \ data. 
These maps, which cover 2442 square-degrees of the Southern sky, represent a significant improvement over previous such products available in this region by virtue of their higher angular resolution ($1\farcm{25}$ and $1\farcm{75}$ for the minimum-variance and component-nulled products, respectively) and lower noise at small angular scales. 
We have detailed the construction of these maps via the linear combination of individual frequency maps from the two experiments including our technique for limiting the correlation of our lowest-variance Compton-$y$ map products with the CIB. 
The new component maps and associated data products, as well as the individual frequency maps from SPT-SZ used to construct these maps, are publicly available at \webaddress \ and the NASA/LAMBDA website.

We have performed several validation tests on the component maps. 
Our analysis of the CMB-CIB-nulled Compton-$y$ map power spectrum shows good agreement with expectations from our adopted sky model. 
A ``blind'' tSZ cluster search using matched-filter techniques demonstrated the expected performance relative to such searches in SPT and \planck \ data alone. 
Finally, a validation test of our tSZ component nulling in the CMB/kSZ maps via stacking such maps at the location of massive SZ-clusters, demonstrated effective removal of the tSZ signal. 

This paper represents the first release of Compton-$y$ and CMB/kSZ component maps from the SPT collaboration. 
We expect these maps---in combination with current surveys such as the Dark Energy Survey \citep{flaugher15} and eROSITA \citep{predehl10, merloni12}; and near-future galaxy surveys like LSST \citep{lsst09}, Euclid \citep{amendola18}, and SPHEREx \citep{dore14}---to provide powerful constraints on both cosmology and the evolution of the gas properties of galaxies, groups, and clusters across cosmic time.
As detailed in this work, both high-resolution and high-frequency measurements are critical for improving the quality of such component maps. 
Such measurements will be available in the coming years by the inclusion of low-noise 95, 150, and 220~GHz data from the ongoing SPT-3G  \citep{benson14} and Advanced ACTPol \citep{henderson16} experiments, and even wider frequency coverage from future CMB experiments \citep{simonsobservatorycollab19,abazajian19}.

\section*{Acknowledgements} 
We thank the anonymous referee for their insightful comments on this manuscript.

The South Pole Telescope program is supported by the National Science Foundation (NSF) through grants PLR-1248097 and OPP-1852617. Partial support is also provided by the NSF Physics Frontier Center grant PHY- 1125897 to the Kavli Institute of Cosmological Physics at the University of Chicago, the Kavli Foundation, and the Gordon and Betty Moore Foundation through grant GBMF\#947 to the University of Chicago.
Argonne National Laboratory's work was supported by the U.S. Department of Energy, Office of Science, Office of High Energy Physics, under contract DE-AC02-06CH11357.
Work at Fermi National Accelerator Laboratory, a DOE-OS, HEP User Facility managed by the Fermi Research Alliance, LLC, was supported under Contract No. DE-AC02- 07CH11359.
The Melbourne authors acknowledge support from the Australian Research Council's Discovery Projects scheme (DP200101068). 
The McGill authors acknowledge funding from the Natural Sciences and Engineering Research Council of Canada, Canadian Institute for Advanced Research, and the Fonds de recherche du Qu\'ebec Nature et technologies. 
The CU Boulder group acknowledges support from NSF AST-0956135.
The Munich group acknowledges the support by the ORIGINS Cluster (funded by the Deutsche Forschungsgemeinschaft (DFG, German Research Foundation) under Germany's Excellence Strategy -- EXC-2094 -- 390783311), the Max-Planck-Gesellschaft Faculty Fellowship Program, and the Ludwig-Maximilians-Universit\"{a}t M\"{u}nchen.
JV acknowledges support from the Sloan Foundation.
This research made use of APLpy, an open-source plotting package for Python \citep{aplpy2012, aplpy2019}. 

\facilities{NSF/US Department of Energy 10m South Pole Telescope (SPT-SZ), \planck, Herschel Space Observatory (SPIRE)}

\bibliography{../../BIBTEX/spt}

\clearpage

\appendix

\section{Data Products}

We provide the following data products and instructions for their use at  \webaddress \  and the NASA/LAMBDA website. 

\begin{enumerate}
\item Minimum Variance Compton-$y$ maps and Null noise maps at  $1\farcm{25}$ resolution. We provide maps in both HEALPix (N$_\textrm{side}$=8192) and  flat sky ($0\farcm{25}$ pixelation) format.
\item CMB-CIB-nulled  and CMB-nulled Compton-$y$ full and half maps at  $1\farcm{75}$ resolution. HEALPix and flat sky format. 
\item tSZ-CIB-nulled and tSZ-nulled CMB/kSZ full and half maps at  $1\farcm{75}$ resolution. HEALPix and flat sky format. 
\item Point source masks and  Dust masks. 
\item SPT-SZ frequency maps, PSDs, bandpasses, and transfer function plus beam estimates for the SPT-SZ maps used to construct the component maps in this work.
\end{enumerate}

\section{Derivation of the Optimal Linear Combination}
\label{appendix:derivation}
In this appendix, we provide an alternate derivation of the result that 
Equations~\ref{eqn:ymap}, \ref{eqn:psi}, and \ref{eqn:sigmapsi} represent, under the assumptions detailed in 
Section~\ref{sec:ymap}, the minimum-variance unbiased estimate of a map 
of signal $S$ with SED $f$ from individual frequency maps. As discussed in that
section, we model the data in observing
frequency $\nu_i$ towards sky direction $\mathbf{n}$ as contributions
from the signal of interest $S$ and noise (including instrumental and astrophysical 
noise and astrophysical signals other than $S$). Neglecting beam and transfer function
for now, we write:
\begin{equation}
T(\nu_i, \mathbf{n}) =  f(\nu_i) \; S(\mathbf{n}) + n(\nu_i, \mathbf{n})
\end{equation}
or, assuming pixelized data and using Roman indices for observing bands and Greek indices for map pixels:
\beq
T_{i \alpha} =  f_i \; S_\alpha + n_{i \alpha}
\eeq
Adapting the derivation of \citet{haehnelt96}, we note that
the most general linear estimator for $S$
using the data from the individual maps is:
\beq 
\bar{S}(\mathbf{n}) = \sum_i \int W_i (\mathbf{n} - \mathbf{n}')
T(\nu_i,\mathbf{n}') d^2 x', 
\eeq
or
\beq 
\label{eqn:estpix}
\bar{S}_\alpha = \sum_{i,\beta} W_{i \alpha \beta} T_{i \beta}.
\eeq
We require that $\bar{S}_\alpha$ be an unbiased estimate of $S_\alpha$:
\beq
\label{eqn:const}
b_\alpha = \langle \bar{S}_\alpha - S_\alpha \rangle = 0,
\eeq
and we wish to minimize the variance
\beq
\sigma^2 = \left \langle \sum_{\alpha,\beta} 
(\bar{S}_\alpha - S_\alpha)(\bar{S}_\beta - S_\beta) \right \rangle.
\eeq
If the noise in each map has zero mean ($\langle n_{i \alpha} \rangle=
0$), then
\beq
b_\alpha = \sum_{i,\beta} W_{i \alpha \beta} \; f_i \; S_\beta - S_\alpha,
\eeq
where 
\beq
\sigma^2 = \sum_{\alpha,\beta} b_\alpha b_\beta + \sum_{i,j,\alpha,\beta,\gamma,\delta} W_{i \alpha \gamma} W_{j \beta \delta} \langle n_{i \gamma} n_{j \delta} \rangle
\eeq 
Using the Lagrange multiplier $\lambda_\alpha$ to enforce the
constraint $b=0$, we minimize the variance by 
taking the functional derivative with respect to
$W_{i \alpha \beta}$ of the quantity
\beq
L=\sigma^2 + \sum_\alpha \lambda_\alpha b_\alpha
\eeq
and setting the result to zero.  Even without solving for
$\lambda_\alpha$, we see that, up to an overall constant, 
the contribution to the optimal estimator $S_\alpha$ from the map 
at each individual frequency is simply the tSZ sensitivity 
weighted by the inverse band-band-pixel-pixel
noise covariance matrix:
\beq
W_{i \alpha \beta} \propto \sum_j \langle n_{i \alpha} n_{j \beta} 
\rangle^{-1} \; f_j,
\eeq
and the proportionality constant is easily obtained by enforcing 
equation \ref{eqn:const}, yielding:
\beq
W_{i \alpha \beta} = \frac{\sum_j \langle n_{i \alpha} n_{j \beta} 
\rangle^{-1} \; f_j}{\sum_{i,j} \langle n_{i \alpha} n_{j \beta} 
\rangle^{-1} \; f_i \; f_j}.
\eeq

If the various sources of noise in the individual maps can be modeled
as Gaussian, random, statistically isotropic fields, then the spherical harmonic 
transform (or two-dimensional Fourier transform, in the flat-sky limit)
of the pixel-pixel-band-band covariance matrix
will be diagonal along the spatial frequency direction:
\beqar
\mathcal{SHT} \left \{ \langle n_{i\alpha} n_{j \beta} \rangle^{-1} \right \} 
&\equiv& \langle n_{i \ell} n_{j  \ell'} \rangle^{-1} \\
\nonumber &=& C^{-1}_{\ell i j} \delta_{\ell \ell'},
\eeqar
where $C_{\ell i j}$ is the spatial power spectrum of the noise in 
band $\nu_i$ if $j=i$ and the cross-spectrum of the noise in bands
$\nu_i$ and $\nu_j$ if $j \neq i$.  In this case we can rewrite 
equation \ref{eqn:estpix} as:
\beq 
\label{eqn:estpix2}
\bar{S}_\ell = \sum_i W_{i \ell} T_{i \ell},
\eeq
where $T_{i \ell}$ is the spherical harmonic transform (SHT) of the
map in band $\nu_i$, $\bar{S}_\ell$ is the SHT of the estimated
Compton-$y$ map, and 
\beq
W_{i \ell} = \frac{\sum_j C^{-1}_{\ell i j} \; f_j}{\sum_{i,j} C^{-1}_{\ell i j} \; f_i \; f_j}.
\label{eqn:wts}
\eeq
Equation~\ref{eqn:psi} is the flat-sky, anisotropic version of Equation~\ref{eqn:wts}, 
including the beam and transfer function in $f$, such that 
$f(\nu_i) \rightarrow f(\nu_i) R(\nu_i,\mathbf{l}$).

\section{Equivalence to Other Formulations}
\label{appendix:equiv}

Equation 3 in \citet{madhavacheril20}, as derived in, e.g., \citet{remazeilles11}, 
gives the following expression for the optimal band weights for one signal component
$a(\nu)$ while explicitly nulling another signal component $a^\prime(\nu)$:
\beq
w_i = \frac{\left ( a^\prime_j C^{-1}_{jk} a^\prime_k \right ) a_j C^{-1}_{ji} - \left ( a_j C^{-1}_{jk} a^\prime_k \right ) a^\prime_j C^{-1}_{ji}}{\left ( a_j C^{-1}_{jk} a_k \right ) \left ( a^\prime_j C^{-1}_{jk} a^\prime_k \right ) - \left ( a_j C^{-1}_{jk} a^\prime_k \right )^2 }.
\label{eqn:psirem}
\eeq
In this expression, summing over repeated indices is assumed, and the $\ell$ subscript has been
omitted. Adopting this convention and, as in the previous appendix, neglecting beam and 
transfer function effects for simplicity, we can write write Equation~\ref{eqn:psimult} (our expression
for multicomponent weights) as:
\beq
\psi_{ij} \equiv \psi_i(\nu_j) = \left [ f^T \mathbf{N}^{-1} f \right ]^{-1}_{ik} \left [ f^T \mathbf{N}^{-1} \right ]_{kj}.
\eeq
(Incidentally, this is immediately recognizable as the solution to the least-squares problem, with $f$ as the
design matrix and $\mathbf{N}$ as the covariance matrix.) For the specific case of two components, we can 
rewrite $f$ and $\psi$ as
\beq
f_{ij} \equiv f_i(\nu_j) =
\left [ \begin{array}{c}
a(\nu_j)  \\
a^\prime(\nu_j)  \\
\end{array} \right ] \equiv
\left [ \begin{array}{c}
a_j  \\
a^\prime_j  \\
\end{array} \right ]
\eeq
\beq
\psi_{ij} \equiv \psi_i(\nu_j) = 
\left [ \begin{array}{c}
w(\nu_j)  \\
w^\prime(\nu_j)  \\
\end{array} \right ] \equiv
\left [ \begin{array}{c}
w_j  \\
w^\prime_j  \\
\end{array} \right ].
\eeq
Also adopting $C$ for the band-band covariance (as opposed to $\mathbf{N}$), we can write $f^T \mathbf{N}^{-1}$ as
\beq
\left [ f^T \mathbf{N}^{-1} \right ]_{ij} \equiv \left [ f^T C^{-1} \right ]_{ij} = 
\left [ \begin{array}{c}
a_k  C^{-1}_{kj} \\
a^\prime_k  C^{-1}_{kj} \\
\end{array} \right ],
\eeq
and write $f^T \mathbf{N}^{-1} f$ as
\beq
\left [ f^T \mathbf{N}^{-1} f \right ]_{ij} \equiv \left [ f^T C^{-1} f \right ]_{ij} = 
\left [ \begin{array}{cc}
a_k C^{-1}_{km} a_m & a_k C^{-1}_{km} a^\prime_m  \\
a^\prime_k C^{-1}_{km} a_m  & a^\prime_k C^{-1}_{km} a^\prime_m  \\
\end{array} \right ],
\label{eqn:nf}
\eeq
Inverting this by hand (and noting that $C$ is symmetric) yields
\beq
\left [ f^T C^{-1} f \right ]^{-1}_{ij} = 
\frac{1}{\left (a_k C^{-1}_{km} a_m \right ) \left (a^\prime_k C^{-1}_{km} a^\prime_m \right ) - \left (a_k C^{-1}_{km} a^\prime_m \right )^2}
\left [ \begin{array}{cc}
a^\prime_m C^{-1}_{km} a^\prime_m & -a_k C^{-1}_{km} a^\prime_m  \\
-a_k C^{-1}_{km} a^\prime_m & a_k C^{-1}_{km} a_m \\
\end{array} \right ],
\eeq
and applying the result to Equation~\ref{eqn:nf} yields
\beq
\left [ \begin{array}{c}
w_j  \\
w^\prime_j  \\
\end{array} \right ] =
\frac{1}{\left (a_k C^{-1}_{km} a_m \right ) \left (a^\prime_k C^{-1}_{km} a^\prime_m \right ) - \left (a_k C^{-1}_{km} a^\prime_m \right )^2}
\left [ \begin{array}{c}
\left ( a^\prime_k C^{-1}_{km} a^\prime_m \right )  a_k C^{-1}_{kj} - \left ( a_k C^{-1}_{km} a^\prime_m \right ) a^\prime_k C^{-1}_{kj}  \\
\left ( a_k C^{-1}_{km} a_m \right ) a^\prime_k C^{-1}_{kj} - \left ( a_k C^{-1}_{km} a^\prime_m \right ) a_k C^{-1}_{jk}  \\
\end{array} \right ],
\eeq
the top row of which is recognizable as Equation~\ref{eqn:psirem} or Equation 3 in \citet{madhavacheril20}.

\end{document}